\documentclass[aps,prb,twocolumn,nobibnotes]{revtex4-2}  
\usepackage{dcolumn}
\usepackage{graphicx}
\usepackage{color}
\usepackage{amssymb}
\usepackage{amsmath}
\usepackage{bm}
\usepackage[colorlinks=true,linktocpage=true,breaklinks=true]{hyperref}
\usepackage{multirow}
\usepackage{braket}
\usepackage{appendix}
\usepackage{mathtools}

\begin{document}

\title{Spin-orbit torque for field-free switching in C$_{\rm 3v}$ crystals}

\author{Diego Garc\'ia Ovalle$^1$}
\email{email: diego-fernando.garcia-ovalle@univ-amu.fr}
\author{Armando Pezo$^1$}
\email{email: armando-arquimedes.pezo-lopez@univ-amu.fr}
\author{Aur\'{e}lien Manchon$^1$}
\email{email: aurelien.manchon@univ-amu.fr}
\affiliation{$^1$Aix-Marseille Universit\'e, CNRS, CINaM, Marseille, France}

\date{\today}

\begin{abstract}
Spin-orbit torques in noncentrosymmetric polycrystalline magnetic heterostructures are usually described in terms of field-like and damping-like torques. However, materials with a lower symmetry point group can exhibit torques whose behavior substantially deviates from the conventional ones. In particular, based on symmetry arguments it was recently proposed that systems belonging to the $C_{\rm 3v}$ point group display spin-orbit torques that can promote field-free switching [Liu et al. Nature Nanotechnology {\bf 16}, 277 (2021)]. In the present work, we analyze the general form of the torques expected in $C_{\rm 3v}$ crystals using the Invariant Theory. We uncover several new components that arise from the coexistence of the three-fold rotation and mirror symmetries. Using both tight binding model and first principles simulations, we show that these unconventional torque components arise from the onset of trigonal warping of the Fermi surface and can be as large as the damping-like torque. In other words, the Fermi surface warping is a key indicator to the onset of field-free switching in low symmetry crystals.
\end{abstract}

\maketitle

\section{Introduction} Electrical manipulation of the magnetization in single magnetic thin films using spin-orbit torques has become routinely available in the past decade \cite{Manchon2019}. In perpendicularly magnetized systems, the most suitable configuration for memory applications, achieving reversible current-driven switching necessitates the combination of spin-orbit torque with an external magnetic field \cite{Miron2010,Liu2011}. As a matter of fact, whereas the spin-orbit torque tends to bring the magnetization in the plane, applying an additional external field along the current direction provides the necessary force that completes the reversal process in a deterministic manner. The need for this external field is considered as a hurdle for memory applications and several strategies have been proposed to circumvent this difficulty. Field-free current-driven switching has been realized using exchange bias from a neighboring antiferromagnet \cite{Oh2016,Fukami2016a}, exchange coupling \cite{Lau2016,Wei2022} or anomalous Hall torque from a proximate ferromagnet \cite{Baek2018,Ryu2022}. The latter takes advantage of an interfacial spin rotation of the incoming spin current \cite{Amin2018}, sometimes called spin swapping \cite{Lifshits2009,Saidaoui2016} (see also Refs. \onlinecite{Luo2019b,Pauyac2018}). In addition, structural engineering has been successfully exploited to design lateral \cite{Yu2014,Bose2018a,Cui2019,Razari2020,Kang2021} and geometrical \cite{Safeer2016} symmetry breaking, tilted anisotropy \cite{You2015,Liu2019f,Li2019l} and longitudinal (compositional or structural) gradient \cite{Shu2022,Chen2019c}. 

Whereas most of these works considered multilayers made out of polycrystalline materials, recent experiments demonstrated that low symmetry crystals are endowed with unconventional spin-orbit torques that can play the role of an external field, thereby completing the current-driven switching process. The impact of the crystalline symmetries on the spin-orbit torque is well-known since its initial observation in the non-centrosymmetric magnetic semiconductors (Ga,Mn)As \cite{Chernyshov2009,Endo2010} and in the Heusler alloy MnNiSb \cite{Ciccarelli2016}, where the bulk inversion symmetry breaking promotes a so-called Dresselhaus-like spin-orbit torque. In fact, further lowering of the crystalline symmetries can lead to unusual torques that turn out to be instrumental to achieve field-free switching. For instance, WTe$_2$ has been shown to display a "perpendicular damping-like torque" \cite{MacNeill2017, MacNeill2017b} that enables field-free switching, an effect confirmed in several experiments \cite{Shi2019,Xie2021,Kao2022}. This torque, also present in MoTe$_2$ \cite{Xue2020} and NbSe$_2$ \cite{Guimaraes2018}, is associated with a crystalline mirror symmetry breaking perpendicular to the interface plane. When a current is injected along this mirror, it may generate a nonequilibrium spin density contained in this mirror plane and normal to the interface. Antiferromagnets are also currently attracting attention from this standpoint. Indeed, the combination of crystalline and magnetic symmetries tend to produce spin currents with a polarization different from what is dictated by the conventional spin Hall effect \cite{Zelezny2017b,Zhang2018d}, an effect sometimes called "magnetic" spin Hall effect \cite{Kimata2019,Ghosh2022}. These spin currents can in turn exert "unconventional" torques on an adjacent ferromagnet, as observed in collinear (Mn$_2$Au \cite{Chen2020c}, RuO$_2$\cite{Bose2022,Bai2022}), and non-collinear antiferromagnets (Mn$_3$GaN \cite{Nan2020}, Mn$_3$Pt \cite{Bai2021} and Mn$_3$Sn \cite{Koudou2021}). 

\begin{table*}[t!]
\centering
\begin{tabular}{cccc|ccc}
& E & $2C_{3}$ & 3$\sigma_v$ &Linear & Quadratic & Cubic\\\hline\hline
\multirow{2}{*}{A$_1$} & \multirow{2}{*}{1} & \multirow{2}{*}{1} & \multirow{2}{*}{1} & $z$ & $x^2+y^2$, $z^2$ & $z^3$, $z(x^2+y^2)$, $x(x^2-3y^2)$\\
&&&& & $m_x^2+m_y^2$, $m_z^2$ & $m_y(3m_x^2-m_y^2)$\\\hline
 
 \multirow{2}{*}{A$_2$} & \multirow{2}{*}{1} & \multirow{2}{*}{1} & \multirow{2}{*}{-1} & &  & $y(3x^2-y^2)$\\
&&&&  $m_z$ & - &$m_x(m_x^2-3m_y^2)$\\\hline
 
 \multirow{3}{*}{E} & \multirow{3}{*}{2} & \multirow{3}{*}{-1} & \multirow{3}{*}{0} & ($x,y$) & $(x^2-y^2,xy)$, $(xz,yz)$ & $(z(x^2-y^2),xyz)$, $(xz^2,yz^2)$, ($x(x^2+y^2),y(x^2+y^2)$)  \\
&&&&  ($m_x,m_y$) & $(m_x^2-m_y^2,m_xm_y)$& $(m_z(m_x^2-m_y^2),m_xm_ym_z)$, $(m_xm_z^2,m_ym_z^2)$\\
&&&&  & $(m_xm_z,m_ym_z)$ & ($m_x(m_x^2+m_y^2),m_y(m_x^2+m_y^2)$)
\end{tabular}\caption{Character table of the C$_{\rm 3v}$ point group. $(x,y,z)$ are the components of a polar vector whereas $(m_x,m_y,m_z)$ are the components of an axial vector.\label{Table1}}
\end{table*}

Recently, \citet{Liu2021} studied the current-driven magnetization reversal in a crystalline CuPt/CoPt bilayer in the L1$_1$ phase grown along the (111) direction. They reported that field-free switching could be achieved when the current was applied along low-symmetry crystallographic directions. Intriguingly, the polarity of the magnetization reversal loop displayed a periodic pattern depending on the crystallographic direction along which the current was applied. This unusual behavior was interpreted as arising from an unconventional torque, tagged "3m" torque, which appears in crystals with C$_{\rm 3v}$ point group \cite{Zelezny2017}. Nonetheless, no microscopic explanation was proposed to explain the emergence of the "3m" torque in this bilayer. Such an explanation is highly desired, especially with the acceleration of the research in two-dimensional van der Waals magnets \cite{Gong2019}. As a matter of fact, most of the van der Waals magnets possess a hexagonal or trigonal point group and are therefore entitled to display such a torque. For instance, the "3m" torque was identified in Fe$_3$GeTe$_2$ monolayer \cite{Johansen2019,Zhang2021} and is associated with an unconventional form of Dzyaloshinskii-Moriya interaction \cite{Laref2020c}. Nonetheless, mere symmetry consideration is not sufficient and a microscopic description is needed. Indeed, recent first principles calculation in the Janus monolayer VSeTe demonstrated that although this material possesses the C$_{\rm 3v}$ symmetry, no "unconventional" torque can be obtained and only the usual field-like and damping-like torques are present \cite{Smaili2021}. Therefore, understanding the physical origin of the "3m" torque in C$_{\rm 3v}$ crystals and suggesting guidelines to enhance it is of crucial interest.

In this work, we intend to clarify the nature of the spin-orbit torque in crystals with C$_{\rm 3v}$ point group, i.e., its vectorial form and its microscopic origin. We first determine the general form of the spin-orbit torque up to the third order in magnetization using the Invariant Theory applied on the C$_{\rm 3v}$ character table. We then consider a minimal model for a magnetic gas with C$_{\rm 3v}$ symmetries. In this model, the spin texture is governed by the cooperation between linear (Rashba) and cubic spin-momentum locking terms. The Fermi surface is characterized by trigonal warping that appears close to the top of the band structure. We show that the unconventional "3m" torque is associated with the cubic spin-momentum locking when the Fermi surface displays strong trigonal warping. We therefore suggest that trigonal warping can be used as a good indicator for the search of "3m" torques in C$_{\rm 3v}$ crystals and two-dimensional van der Waals magnets.

\section{Symmetry analysis} We first determine the general form of the torque using the representation theory \cite{Dresselhaus2008,Lax2012}. The C$_{\rm 3v}$ point group is characterized by the identity $E$, the three-fold rotation along {\bf z}, C$_3$, and the mirror symmetry normal to, say, {\bf y}, $\sigma_v$. It has three irreducible representations $A_1$, $A_2$ and $E$, i.e., matrices {\it representing} the action of the symmetry operations $E$, C$_3$ and $\sigma_v$. Although a given symmetry operation can be represented by an infinite number of matrices, the trace of these representative matrices is unique for a given operation. Therefore, each irreducible representation can be identified by a unique set of traces called "characters". Table \ref{Table1} gives the character table of the C$_{\rm 3v}$ point group. The (equilibrium and nonequilibrium) properties of a given crystal are written as the combination of polar and axial vectors. For instance, in the case of the spin-orbit torque these vectors are the electric field $(E_x, E_y, E_z)$ (polar vector) and the magnetization $(m_x, m_y, m_z)$ (axial vector). Concretely, they transform in the following way,
\begin{eqnarray}\label{eq:polar}
(E_x,E_y)&\xrightarrow{\sigma_v}&(E_x,-E_y),\\
(E_x,E_y)&\xrightarrow{C_3}&(-\tfrac{1}{2}E_x-\tfrac{\sqrt{3}}{2}E_y,\tfrac{\sqrt{3}}{2}E_x-\tfrac{1}{2}E_y),
\end{eqnarray}
and
\begin{eqnarray}\label{eq:axial}
(m_x,m_y,m_z)&\xrightarrow{\sigma_v}&(-m_x,m_y,-m_z),\\
(m_x,m_y,m_z)&\xrightarrow{C_3}&(-\tfrac{1}{2}m_x-\tfrac{\sqrt{3}}{2}m_y,\tfrac{\sqrt{3}}{2}m_x-\tfrac{1}{2}m_y,m_z).\nonumber\\
\end{eqnarray}

When applying the symmetry operations on these vectors' components, they transform according to the irreducible representations $A_1$, $A_2$ and $E$ so that one can define {\it basis functions} for each representation. In Table \ref{Table1}, we give the basis functions of the irreducible representation of the C$_{\rm 3v}$ point group up to the third order in magnetization.

Let us determine the general form of the torque based on Table \ref{Table1}. The spin-orbit torque ${\bm \tau}$ is associated with an effective field, ${\bm\tau}=-\gamma {\bf m}\times {\bf h}$. This effective field ${\bf h}$ is an axial vector, like {\bf m}, so its component $h_z$ and $(h_x,\;h_y)$ belong to $A_2$ and $E$, respectively, and can be expanded as combinations of the invariant basis functions given in Table \ref{Table1}. The only possible combinations of magnetization components with $\textbf{E}$ that are invariant under the symmetries of the group are $m_z$, $m_z^3$ and $m_x(m_x^2-3m_y^2)$, that all belong to $A_2$. Conversely, $\textbf{z}\times\textbf{E}$ is an axial vector, and hence its allowed combinations are $1$, $m_z$ and $m_y(3m_x^2-m_y^2)$, that all belong to $A_1$. Accounting for all combinations involving polar vector components at the first order and axial vector components up to the third order in magnetization, we obtain

\begin{widetext}
\begin{eqnarray}\label{eq:1}
{\bm h}_{\|}=&& h_{\rm FL}^{{\|}}(1+\eta_{\rm FL} m_z^2+\delta_{\rm FL}m_y(3m_x^2-m_y^2)){\bf z}\times{\bf E}+h_{\rm DL}^{{\|}}((1+\eta_{\rm DL} m_z^2)m_z+\delta_{\rm DL} m_x(m_x^2-3m_y^2)){\bf E}\\\nonumber
&&+[h_{\rm 3m}^{{\|}}m_x(1+\eta_{\rm 3m} m_z^2)+h_{\rm 3m}^{\rm z} m_zm_y-2h_{\rm PH}^{{\|}}m_xm_y+h_{\chi}^{{\|}}m_z(m_x^2-m_y^2)](E_x{\bf x}-E_y{\bf y})\\\nonumber
&&+[-h_{\rm 3m}^{{\|}}m_y(1+\eta_{\rm 3m} m_z^2)+h_{\rm 3m}^{\rm z}m_zm_x+h_{\rm PH}^{{\|}}(m_x^2-m_y^2)+2h_{\chi}^{{\|}} m_xm_ym_z] (E_y{\bf x}+E_x{\bf y}),\\
\label{eq:2}
{\bm h}_{\bot}=&& \left[h_{\rm DL}^{\rm z}(1+\eta_{\rm z} m_z^2){\bf E}\cdot{\bf m}+h_{\rm FL}^{\rm z}m_z{\bf m}\cdot({\bf z}\times{\bf E})\right.\\
&&\left.+h_{\rm PH}^{\rm z} ((m_x^2-m_y^2)E_y+2m_xm_yE_x)+h_{\chi}^{\rm z}m_z ((m_x^2-m_y^2)E_x-2m_xm_yE_y)\right]{\bf z}.\nonumber
\end{eqnarray}
\end{widetext}

The formulas given above are general and they do not rely on any specific microscopic mechanism. We recognize the field-like torque ($h_{\rm FL}^{{^{\|,\rm z}}}$), the damping-like torque ($h_{\rm DL}^{{^{\|,\rm z}}}$) and the "3m" torque reported in Refs. \onlinecite{Liu2021,Johansen2019} ($h_{\rm 3m}^{{^{\|,\rm z}}}$). At higher orders, these terms are modulated by a planar anisotropy term, $\sim \eta_\alpha$, and a trigonal anisotropy term, $\sim \delta_\alpha$. In addition, the magnitude of the field-like and damping-like torques are different in-plane ($h_{\rm FL}^{{\|}}$, $h_{\rm DL}^{{\|}}$) and out-of-plane ($h_{\rm FL}^{\rm z}$, $h_{\rm DL}^{\rm z}$). By removing these anisotropies, i.e., by setting $\eta_{\rm FL,DL}=0$, $\delta_{\rm FL,DL}=0$ and $h_{\rm FL,DL}^{{\|}}=h^z_{\rm FL,DL}$, one retrieves the effective fields associated with the conventional field-like and damping-like torques, i.e., $\sim{\bf z}\times{\bf E}$ and $\sim({\bf z}\times{\bf E})\times{\bf m}$.\par

In addition, we also identify two additional torques that we refer to as in-plane ($h_{\rm PH}^{{\|}}$) and out-of-plane ($h_{\rm PH}^{\rm z}$) planar Hall torque, and chiral torques ($h_{\chi}^{{\|,\rm z}}$). The planar Hall torque possesses symmetries comparable to the planar Hall effect: it is active when the magnetization lies in the $(\bf{x}, \bf{y})$ plane and its magnitude depends on the angle between the electric field and the magnetization. The chiral torque necessitates to cant the magnetization away from the plane and it changes sign when reversing the magnetization ($m_z\rightarrow - m_z$).\par

\begin{figure}[ht!]
\includegraphics[width=0.9\linewidth]{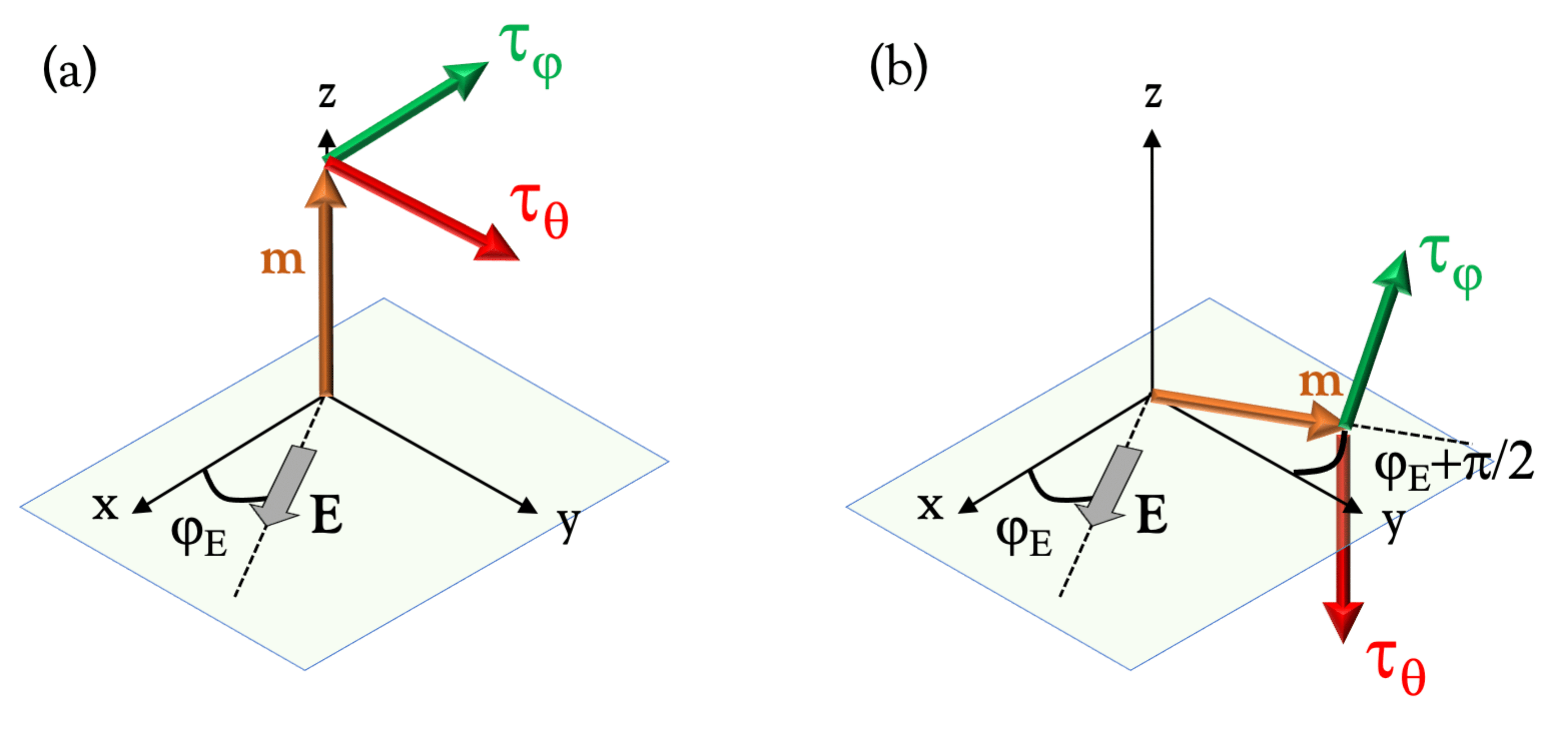}
\caption{(Color online) Schematics of the torque components when the magnetization is (a) perpendicular to the plane and (b) in the plane, at 90° of the applied electric field.\label{Fig1}} 
\end{figure}

To clarify the impact of the torque on the magnetization dynamics, we analyze its expression in two illustrative situations. When the magnetization lies out-of-plane ($\textbf{m}={\bf z}$), which is typical of perpendicularly magnetized systems at rest [see Fig. \ref{Fig1}(a)], the two torque components up to first order in magnetization read

\begin{align}
\tau_{\|}&=-\gamma h_{\rm FL}^{\|}\textbf{z}\times (\textbf{z}\times \textbf{E}),\label{eq:t1}\\
\tau_{\perp}&=-\gamma h_{\rm DL}^{\|}\textbf{z}\times \textbf{E}\label{eq:t2}
\end{align}

We see that only the conventional field-like and damping-like torques are active in this configuration. One can see that the field-like torque is always along the electric field, $\sim {\bf m}\times({\bf z}\times {\bf E})$, whereas the damping-like torque is perpendicular to it $\sim {\bf m}\times[({\bf z}\times {\bf E})\times{\bf m}]$. These two torques are the ones that destabilize the magnetization from its rest position and tend to bring it in the plane, normal to the applied electric field [see Fig. \ref{Fig1}(b)].



Once the magnetization is in-plane, at $\phi=\varphi_E+\frac{\pi}{2}$, where $\varphi_E$ is the in-plane angle of the electric field with respect to {\bf x} and $(\theta,\phi)$ are the polar and azimuthal angles of the magnetization unit vector, the torques ${\bm \tau}=-\gamma{\bf m}\times {\bf h}={\bf \tau}_\theta{\bf e}_\theta+{\bf \tau}_\phi{\bf e}_\phi$ in regular spherical coordinates are

\begin{align}
\tau_\theta/E&=\gamma[h_{\rm DL}\delta_{\rm DL}-h_{3m}]\sin 3\varphi_E,\\
\tau_\phi/E&=\gamma h_{\rm PH}^z\cos 3\varphi_E.
\end{align}

In this configuration, the conventional field-like and damping-like torques are quenched, and the only active torques are the "3m" torque ($h_{\rm 3m}^{{\|}}$), identified in Ref. \onlinecite{Liu2021}, the trigonal anisotropy correction to the damping-like torque ($h_{\rm DL}^{{\|}}\delta_{\rm DL}$), and the perpendicular planar Hall torque ($h_{\rm PH}^z$). Here, only $\tau_\theta$ induces the deterministic switching, which means that the "3m" torque and the trigonal anisotropy correction to the damping-like torque are the active contributions in this process. Remarkably, in this frame the two other torques identified in Eqs. \eqref{eq:1}-\eqref{eq:2}, i.e. the planar Hall torque ($h_{\rm PH}^{{\|,\rm z}}$) and the chiral torque ($h_{\chi}^{{\|,\rm z}}$), are only active when $\theta\neq0,\pi/2$ and should therefore impact the magnetization dynamics itself. Their influence could modify the current-driven auto-oscillation \cite{Liu2012b,Demidov2012}, a phenomenon that we leave to future studies.

\section{Physical origin of the unconventional torques}
\subsection{Minimal tight-binding model for $C_{3v}$ magnets} 
\begin{figure}[ht!]
\includegraphics[width=0.9\linewidth]{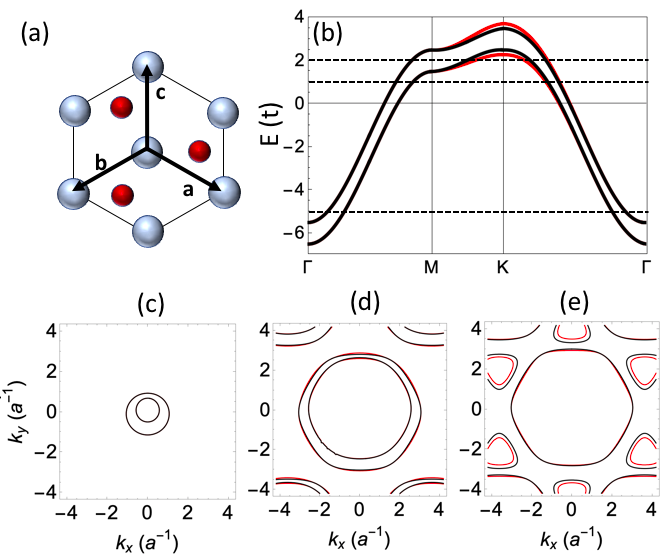}
\caption{(Color online)(a) The unit cell of the minimal model for the C$_{3v}$ crystal. The grey atoms represent the hexagonal lattice sites and the red atoms break the plane inversion symmetry while conserving the three-fold rotation along {\bf z} and the mirror symmetry normal to {\bf y}. (b) Band structure of the tight-binding model described in the text with $t_R=0.1t$ and $\Delta=0.5t$, for the cases $t_{R3}=0$ (black lines), and $t_{R3}=0.1t$ (red lines). The magnetization direction is set to $\theta=\frac{\pi}{2}$ and $\phi=0$. The horizontal dashed lines correspond to $\mu=-5t$,  $\mu=t$ and $\mu=2t$, respectively. (c-e) Fermi surfaces with ($t_{R3}=0.1t$ - red lines) and without ($t_{R3}=0$ - black lines) cubic spin-orbit coupling for (c) $\mu=-5t$, (d) $\mu=t$ and (e) $\mu=2t$.}\label{Fig2} 
\end{figure}

The symmetry analysis provided above does not give information about the relative magnitude of the different torques. To better understand which microscopic mechanisms control these different components, we now turn our attention towards a minimal model for the spin-orbit torque. We consider a ferromagnetic system defined in a hexagonal lattice as depicted on Fig. \ref{Fig2}(a) with C$_{\rm 3v}$ symmetry modeled by the Hamiltonian 
\begin{eqnarray}
{\cal H}_0=\varepsilon_{\bf k}+\Delta{\bm\sigma}\cdot{\bf m}+{\cal H}_{\rm R}+{\cal H}_{\rm R3}\label{7},
\end{eqnarray}
with 
\begin{eqnarray}
&&{\cal H}_{\rm R}=-i\frac{t_{\rm R}}{a}\sum_{{\bf u},s=\pm} s{\bm\sigma}\cdot({\bf z\times u})e^{is{\bf k}\cdot{\bf u}}=\frac{t_{\rm R}}{a}{\bm \eta}_{\bf k}\cdot({\bm\sigma}\times{\bf z}),\nonumber\\
\\
&&{\cal H}_{\rm R3}=-it_{\rm R3}\sum_{{\bf u},s=\pm} s{\bm\sigma}\cdot{\bf z}e^{is{\bf k}\cdot{\bf u}}=t_{\rm R3}{\bm\lambda}_{\bf k}\sigma_z.
\end{eqnarray}
The sum is taken over the nearest neighbors, i.e., ${\bf u}={\bf a},{\bf b},{\bf c}$, sketched on Fig. \ref{Fig2}(a), and $a$ is the lattice parameter. Explicitly, $\varepsilon_{\bf k}=-2t(\cos{\bf k}\cdot{\bf a}+\cos{\bf k}\cdot{\bf b}+\cos{\bf k}\cdot{\bf c})$, ${\bm \eta}_{\bf k}=2({\bf a}\sin{\bf k}\cdot{\bf a}+{\bf b}\sin{\bf k}\cdot{\bf b}+{\bf c}\sin{\bf k}\cdot{\bf c})$ and ${\bm\lambda}_{\bf k}=2(\sin{\bf k}\cdot{\bf a}+\sin{\bf k}\cdot{\bf b}+\sin{\bf k}\cdot{\bf c})$. Here, $t$ is the nearest-neighbor hopping parameter, $\Delta$ is the exchange between the conduction electrons and the magnetization ${\bf m}$, $t_R$ is the linear Rashba spin-orbit coupling coming from inversion symmetry breaking normal to the ({\bf a}, {\bf b}) plane and $t_{\rm R3}$ is its cubic correction that is associated with the mirror symmetry normal to {\bf y} axis.

\noindent 
This model enhances the conventional free-electron Rashba gas by adding two ingredients: (i) the hexagonal symmetry and (ii) the cubic spin-orbit coupling. Figure \ref{Fig2}(b) represents the band structure for a standard set of parameters with (red lines) and without (black lines) cubic spin-orbit coupling. One directly sees that the cubic spin-orbit coupling only modifies the band structure close to the top of the band. The Fermi surface at three characteristic fillings are sketched on Figs. \ref{Fig2}(c-e), with (red lines) and without (black lines) cubic spin-orbit coupling. At low band filling [Fig. \ref{Fig2}(c)], where the dispersion is mostly quadratic, the Fermi surface is spherical and the cubic spin-orbit coupling has almost no impact. We therefore expect that the torque reduces to its most conventional form, field-like and damping-like. Upon increasing the band filling [Fig. \ref{Fig2}(d,e)], the Fermi surface starts displaying hexagonal warping and the cubic spin-orbit coupling modifies the energy contours. In this context, at high band filling the warping is strong with Fermi pockets appearing away from $\Gamma$-point. Turning on the cubic spin-orbit coupling modifies the Fermi surface, resulting in a strong trigonal warping. It is clear that the unconventional torques identified in the previous section are expected to emerge in this regime.\par

The impact of Fermi surface warping on the spin-orbit torque has been addressed theoretically in the context of the topological insulator surfaces. \citet{Kurebayashi2019,Imai2021} investigated the influence of warping on the spin-transfer torque and spin-orbit torque, respectively, in magnetic domain walls and skyrmions to the first order of the magnetization gradient. The spin-orbit torque discussed presently is not addressed in these works. \citet{Zhou2022} investigated the appearance of a damping-like torque that is nonlinear in electric field and directly induced by the warping. \citet{Li2019k} investigated the impact of the hexagonal warping on the spin-orbit torque, linear in electric field, and observed that the torque does not vanish when the magnetization lies in the plane. This is consistent with the analysis performed in the previous section, although a direct connection with the general form provided in Eqs. (\ref{eq:1})-(\ref{eq:2}) remains difficult. \par 

Let us now compute the effective spin-orbit field driven by the current, ${\bf h}=(\Delta/V M_s)\braket{\bm \sigma}$, where $M_s$ is the saturation magnetization of the ferromagnet, $V$ is the volume of the unit cell and  $\braket{...}$ denotes nonequilibrium quantum statistical averaging. $\braket{\bm \sigma}$ is computed within the linear response formalism considering the symmetrized decomposition of Kubo-Bastin formula proposed in Ref. \cite{Bonbien2020}, which takes the form
\begin{align}
  &\braket{\hat{\sigma}_i}_{\rm Int}=-\frac{e\hbar}{4\pi}\int f(\epsilon)d\epsilon \operatorname{Re}\left[\operatorname{Tr}\left\{\hat{\textbf{v}}(G^{\rm R-A})\hat{\sigma}_i(\partial_\epsilon G^{\rm R+A})\right\}\right],\label{8}\\ 
  & \braket{\hat{\sigma}_i}_{\rm Ext}=-\frac{e\hbar}{8\pi}\int \partial_\epsilon f(\epsilon)d\epsilon \operatorname{Re}\left[\operatorname{Tr}\left\{\hat{\textbf{v}}(G^{\rm R-A})\hat{\sigma}_i(G^{\rm R-A})\right\}\right].\label{9}
\end{align}
\noindent 
Here $\hat{\textbf{v}}=\partial_{\bf k} \mathcal{H}$ is the velocity operator in the direction of the applied electric field, $f(\epsilon)$ is the equilibrium Fermi distribution function, $G^{\rm R(A)}$ is the retarded (advanced) Green function and $G^{\rm R\pm A}=G^{\rm R}\pm G^{\rm A}$. Notice that Eq. \eqref{8} gives the intrinsic contribution whereas Eq. \eqref{9} gives the extrinsic one. Based on time-reversal symmetry, the spin-orbit field components that are {\em odd} in magnetization are of intrinsic origin, such as the dampinglike torque $h_{\rm DL}^{{\|,z}}$, the "3m" torque $h_{\rm 3m}^{{\|}}$ and the chiral torque $h_{\chi}^{{\|,z}}$, whereas the terms that are {\em even} in magnetization are extrinsic, such as the field-like torque $h_{\rm FL}^{{\|,z}}$ and the planar Hall torque $h_{\rm PH}^{{\|,z}}$.

\begin{figure}[ht!]
\includegraphics[width=1\linewidth]{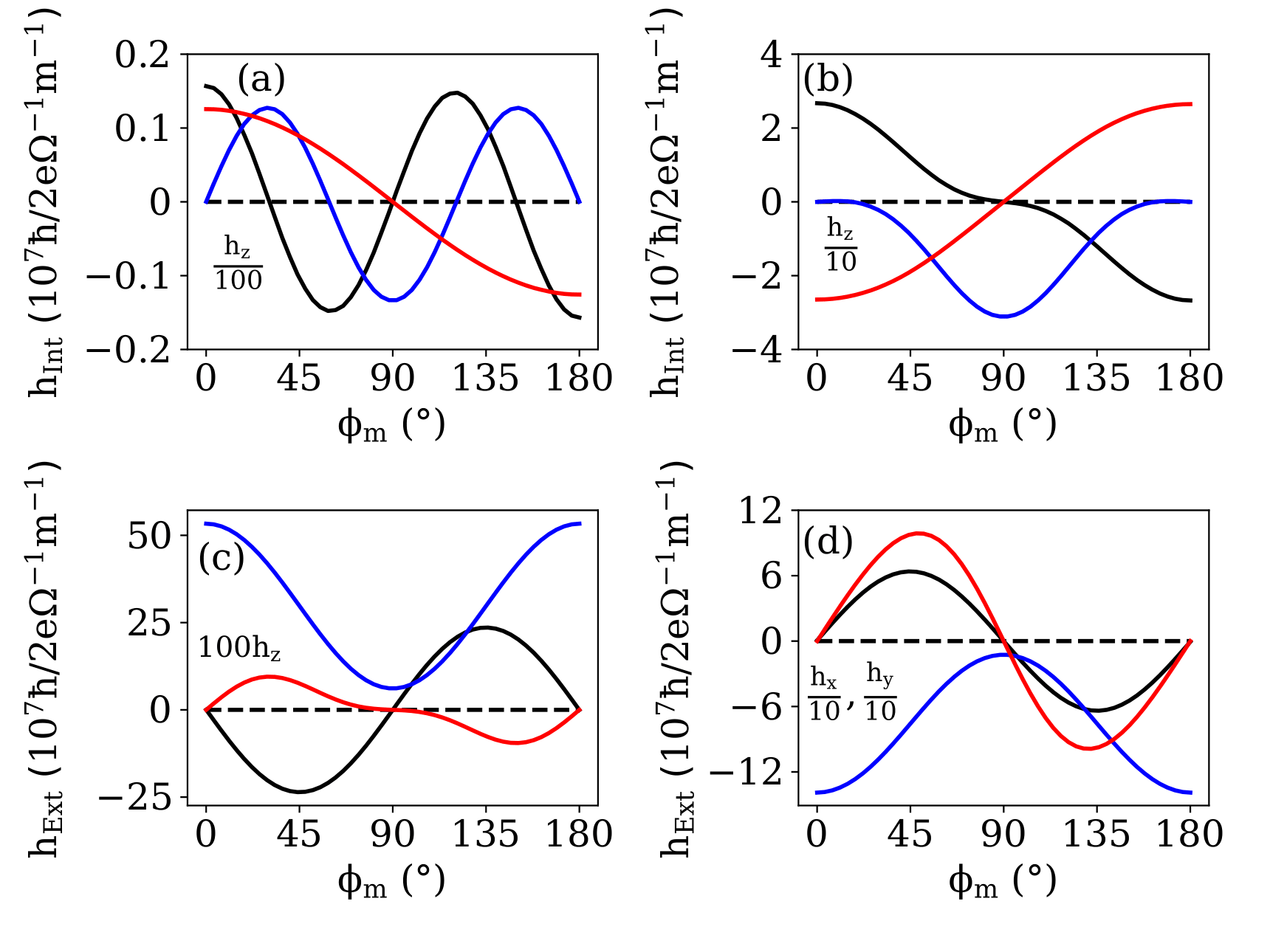}
\caption{(Color online) Angular dependence of the effective field components $h_{x}$ (black lines), $h_y$ (blue lines) and $h_z$ (red lines) when the magnetization rotates in the ($x$, $y$) plane. We indicate a scaling factor of a given component whenever is necessary. The system's parameters are $t=1$, $t_{\rm R}=0.1t$, $t_{\rm R3}=0.05t$, $\Delta=0.5t$ and the homogeneous disorder $\Gamma=0.1t$. The intrinsic (a, b) and extrinsic (c, d) contributions are plotted for $\mu=-5t$ (left panels) and $\mu=2t$ (right panels). Our calculations reproduce our symmetry predictions in Eq. \eqref{hinp}.}
\label{Fig3}
\end{figure}

To assess the relative magnitude of the different torque components and identify their physical origin, we compute the angular dependence of the fields when the magnetization rotates in the (\textbf{x}, \textbf{y}) plane, while the electric field is applied along $x$. Our results are reported in Fig. \ref{Fig3} for both intrinsic (a,b) and extrinsic (c,d) contributions in the cases of low (left panels) and high(right panels) band fillings. We analyze these results based on the angular dependence of the spin-orbit fields given by Eqs. \eqref{eq:1}-\eqref{eq:2}, when there is linear and cubic spin-orbit coupling. In this scenario, we deduce that
\begin{eqnarray}
{\bf h}=\left(\begin{matrix}
h_{\rm 3m}^{{\|}}\cos\phi-h_{\rm PH}^{{\|}}\sin 2\phi+h_{\rm DL}^{{\|}}\delta_{\rm DL}\cos 3\phi\\
-h_{\rm 3m}^{{\|}}\sin\phi+h_{\rm PH}^{{\|}}\cos 2\phi + h_{\rm FL}^{{\|}}(1+\delta_{\rm FL}\sin3\phi)\\
h_{\rm DL}^z\cos\phi+h^z_{\rm PH}\sin2\phi
\end{matrix}\right)\nonumber\\\label{hinp}
\end{eqnarray}
From Fig. \ref{Fig3}(a) it is clear that for $h_x$ and $h_y$ a 3-fold dependence related to $\delta_{\rm DL}$ and $\delta_{\rm FL}$ dominates at low band filling, while for $h_z$ the term $h_{\rm DL}^{\rm z}$ is predominant in this regime. Besides, $h_{\rm 3m}^{{\|}}$ can be identified through the $h_y$ contribution at high band filling [Fig. \ref{Fig3}(b)]. Regarding the extrinsic contributions, we notice from Fig. \ref{Fig3}(c) that $h_x$ and $h_y$ are not trivial due to $h_{\rm PH}^{{\|}}$ and $h_{\rm FL}^{{\|}}$ at low band filling, whereas $h_{\rm PH}^{\rm z}$ becomes relevant at high band filling in Fig. \ref{Fig3}(d).

\begin{figure}[ht!]
\includegraphics[width=1.05\linewidth]{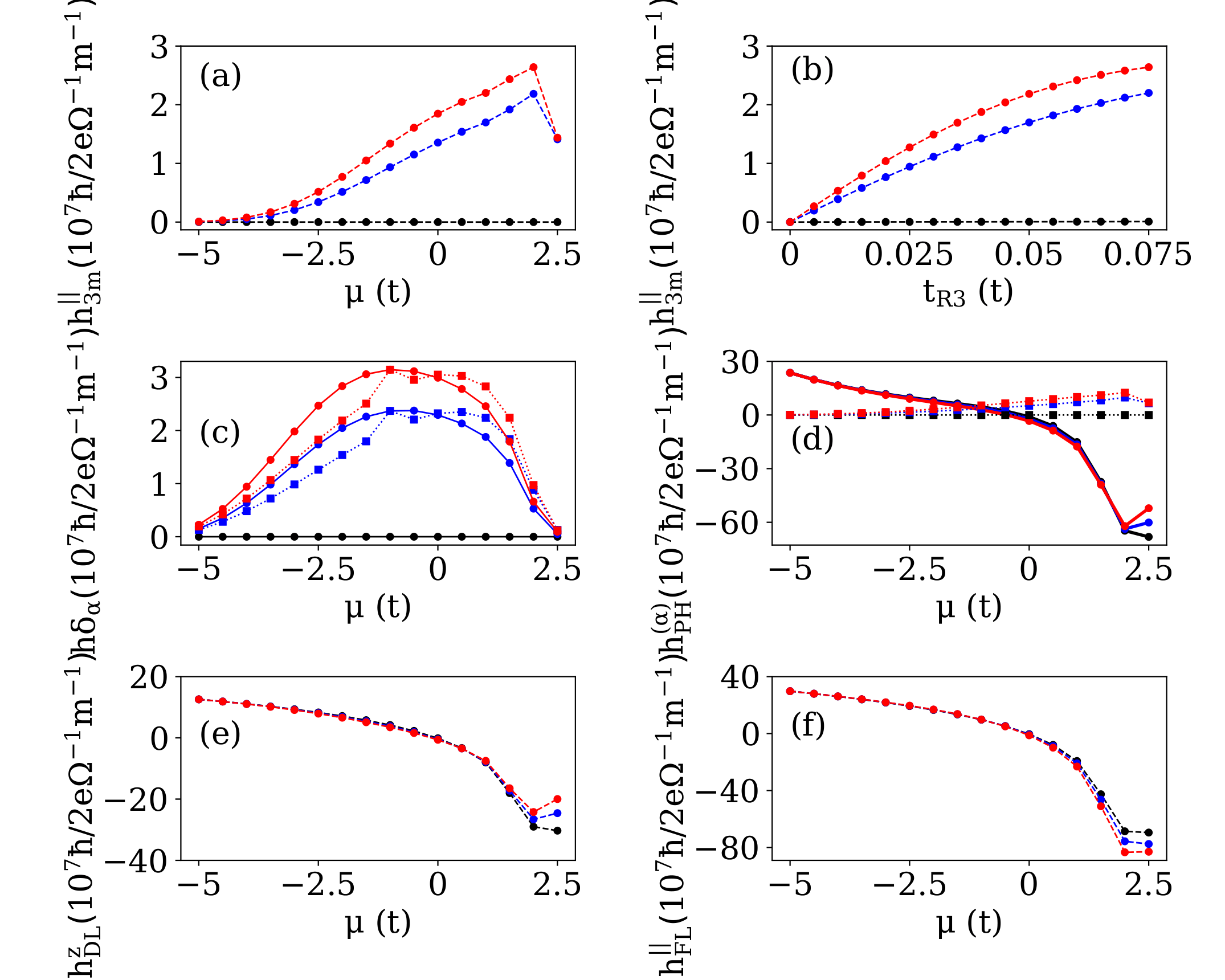}
\caption{(Color online) Effective field's components extracted from fitting the numerical results with Eq. \eqref{hinp}. The panels show (a) $h_{\rm 3m}^{{\|}}$, (c) $h_{\rm DL}^{{\|}}\delta_{\rm DL}$ (solid lines) and $h_{\rm FL}^{{\|}}\delta_{\rm FL}$ (dotted lines), (d) $h_{\rm PH}^{{\|}}$ (solid lines) and $h_{\rm PH}^{\rm z}$ (dotted lines), (e) $h_{\rm DL}^{\rm z}$, and (f) $h_{\rm FL}^{{\|}}$ as a function of the chemical potential $\mu$, for different values of the cubic spin-orbit coupling $t_{\rm R3}$ (black, blue and red lines stand for $t_{R3}=0,\; 0.05t,\;0.075t$). For completeness, panel (b) shows $h_{\rm 3m}^{{\|}}$ as a function of $t_{\rm R3}$ for different values of $\mu$ (black, blue and red lines stand for $\mu=-5t,\;t, \; 2t$).}\label{Fig4}
\end{figure}

Analyzing the in-plane angular dependence of the field components, Fig. \ref{Fig3}, with Eq. \eqref{hinp}, one can extract the different torque contributions, reported on Fig. \ref{Fig4} as a function of $t_{R3}$ and $\mu$. From Figs. \ref{Fig4}(a, b), $h_{\rm 3m}^{{\|}}$ requires cubic Rashba coupling and increases with the band filling, confirming its sensitivity to the trigonal warping of the Fermi surface. This behaviour is different from $h_{\alpha}^{{\|}}\delta_\alpha$ ($\alpha$=DL, FL), which is displayed in Fig. \ref{Fig4}(c) and reaches a maximum close to $\mu=0$ for $t_{R3}\not=0$. The planar contributions $h_{\rm PH}^{{\|}}$ and $h_{\rm PH}^{\rm z}$ are depicted in Figs. \ref{Fig4}(d) and they exhibit different behaviors with $t_{R3}$ and $\mu$: whereas $h_{\rm PH}$ does not require $t_{\rm R3}$ and it follows a similar tendency to $h_{\rm DL}^{\rm z}$and $h_{\rm FL}^{{\|}}$ [Fig. \ref{Fig4}(e, f)], $h_{\rm PH}^{\rm z}$ increases with the band filling and requires $t_{\rm R3}\not=0$. The salient features of the different torque components in C$_{\rm 3v}$ systems are summarized in Table \ref{Table2}.

\begin{table}[ht!]
\centering
\begin{tabular}{c|cc}
Component & Physical origin & Source\\\hline\hline
$h_{\rm FL}^{{\|}}$,\;$h_{\rm FL}^{\rm z}$,\;$h_{\rm PH}^{{\|}}$ & Extrinsic & Linear Rashba\\
$h_{\rm PH}^{\rm z}$,\;$h_{3m}^z$ & Extrinsic & Linear + cubic Rashba\\
$h_{\rm DL}^{{\|}}$,\;$h_{\rm DL}^{\rm z}$ & Intrinsic & Linear Rashba\\
{$\delta_{\rm FL}$},\;$\delta_{\rm DL}$ & Intrinsic & Linear + cubic Rashba\\
$h_{3m}^{{\|}}$,\;$h_{\chi}^{{\|}}$,\;$h_{\chi}^z$ & Intrinsic &  Linear + cubic Rashba\\
\end{tabular}\caption{Summary of the minimal model analysis.\label{Table2}}
\end{table}

\subsection{First principles case study: CuPt(111)/Co} 
\begin{figure}[ht!]
\includegraphics[width=\linewidth]{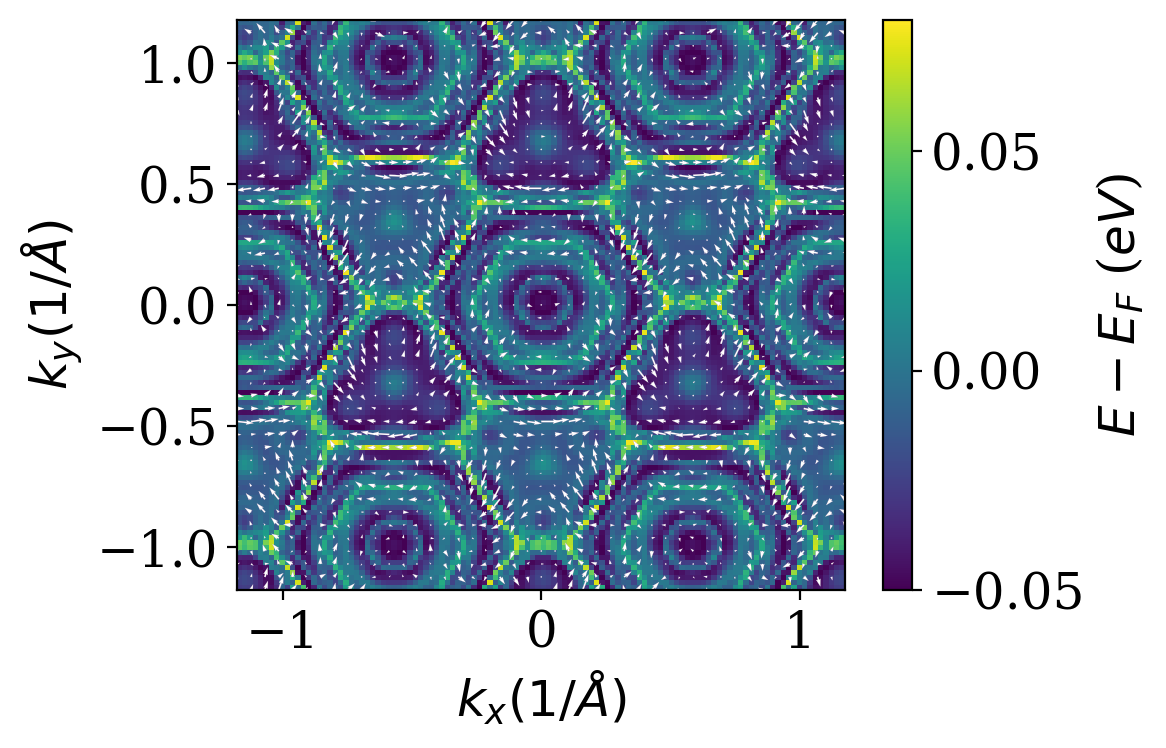}
\caption{(Color online) Spin texture in momentum space close to Fermi level a selected band of CuPt(111)/Co slab computed from first principles. A strong hexagonal symmetry is obtained suggesting the presence of a large cubic spin-orbit coupling interaction.\label{Fig5}} 
\end{figure}

We conclude this work by computing the spin-orbit torque in L1$_1$ CuPt(111)/Co from first principles. As explained above, this material has been recently experimentally demonstrated to host a sizable "3m" torque \cite{Liu2021}. We considered a CuPt/Co slab containing 12 layers, such that the L1$_1$ phase is made up of stacking elemental fcc layers along the [111] direction. We determine the band structure and spin textures by employing fully relativistic density functional theory. We describe the spin-orbit coupling within a fully relativistic pseudo-potential formulation and used the generalized gradient approximation (GGA) for the exchange-correlation functional, the calculations are converged for a 400 Ry plane-wave cut-off for the real-space grid with a 13$\times$13$\times$1 $k$-points sampling of the Brillouin zone. We used the conjugate gradient algorithm to minimize the atomic forces below 0.01 eV/\AA. The momentum-resolved spin texture at the Fermi level is reported in Fig. \ref{Fig5} and displays a very clear hexagonal symmetry, suggesting an effectively large cubic spin-orbit coupling interaction.

\begin{figure}[ht!]
\includegraphics[width=\linewidth]{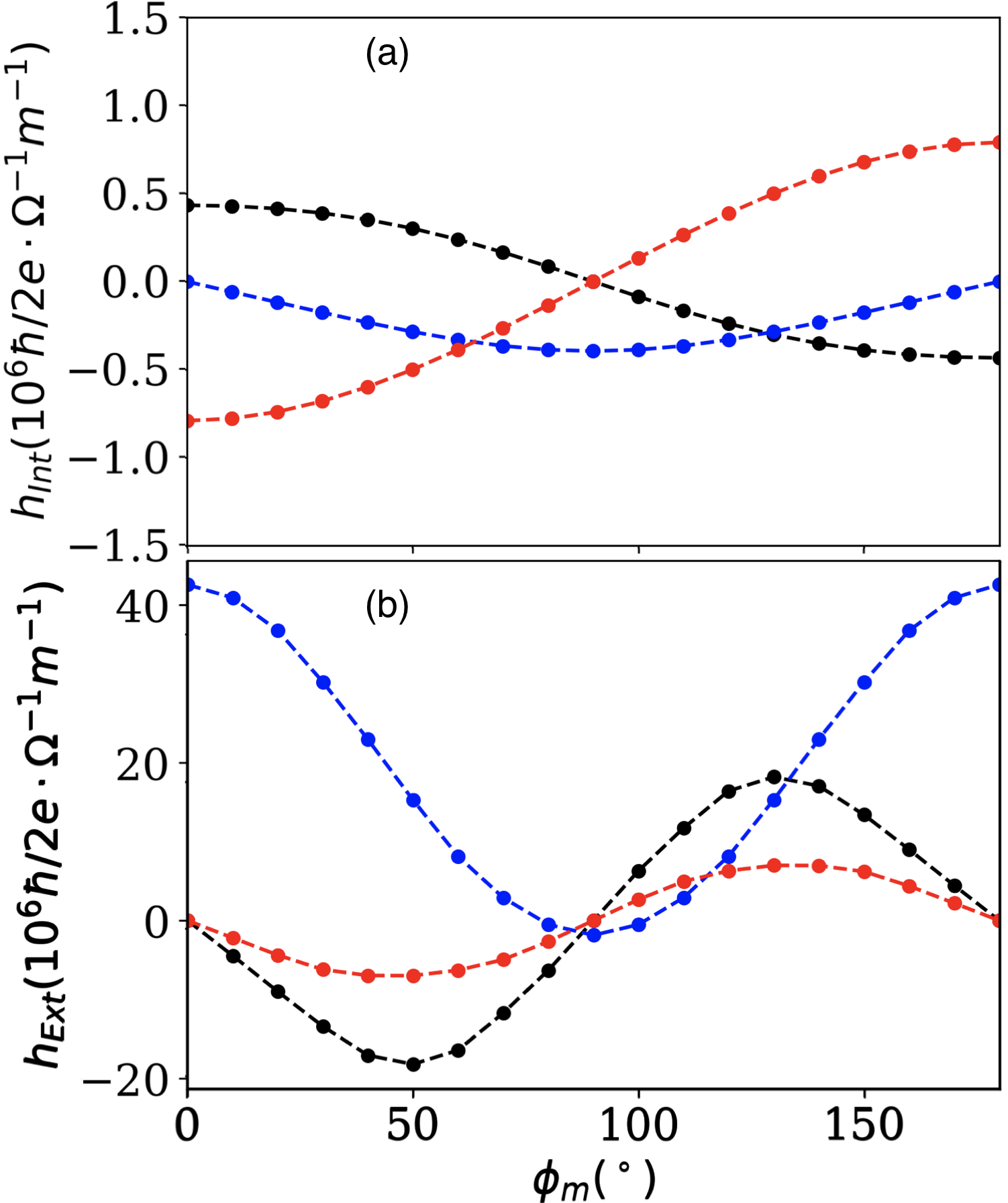}
\caption{(Color online) Angular dependence of the intrinsic (top) and extrinsic (bottom) spin-orbit field components when the magnetization is rotated in the ($x$, $y$) plane. The black, blue and red curves represent the x, y and z components the effective fields, respectively}.\label{Fig6} 
\end{figure}

The angular dependence of the intrinsic and extrinsic spin-orbit fields is reported on Fig. \ref{Fig6}(a) and (b), respectively, when the magnetization is rotated in the ($x$,$y$) plane. The calculations are performed with broadening $\Gamma=0.025$ eV in the zero temperature limit. The angular dependence is well reproduced by Eq. \eqref{hinp}. The intrinsic spin-orbit torque is composed of the damping-like torque ($h^z$) and the "3m" torque  ($h^{x,y}$), with $h_{\rm 3m}^{{\|}}/h_{\rm DL}^z\approx 0.67$, indicating that the "3m" torque is about the same order of magnitude as the damping-like torque. The extrinsic torque is one order of magnitude larger and is composed of the field-like torque and the planar Hall torque. The possible differences between our numerical predictions and our symmetry analysis in Eq. \eqref{hinp} can be explained by the neglect of higher-order terms in the character table expansion and the large values of cubic spin-orbit coupling. Nevertheless, we can extract $h_{\rm PH}^{{\|}}/h_{\rm FL}^{{\|}}\approx 1$ and $h_{\rm PH}^{z}/h_{\rm PH}^{{\|}}\approx 0.4$, meaning that the planar Hall torque is anisotropic and as large as the fieldlike torque, and shall therefore impact the magnetization switching and dynamics. We leave this question to further studies. We emphasize that the relative magnitude of the intrinsic to extrinsic torques is not meaningful since the extrinsic torque is inversely proportional to the disorder broadening $\Gamma$, which is taken as (small) free parameter in our model.

\subsection{Discussion and conclusion} The presence of these unconventional torques is particularly interesting for applications as they not only enable field-free switching but also impact the current-driven auto-oscillations. Our minimal model suggests that C$_{3v}$ crystals could host such torques. Nonetheless, we emphasize that this is not a sufficient condition. As a matter of fact, in a previous study, we computed the spin-orbit torque in vanadium-based Janus transition metal dichalcogenides VSeTe and found no such torque, in spite of the similar crystal symmetries \cite{Smaili2021}. We attributed this absence to the fact that in this material, the electronic transport here is mostly driven by states at $\Gamma$-point and therefore the crystal symmetries are not {\em imprinted} on the Bloch states. In contrast, in the L1$_1$ CuPt the Fermi surface shows a very strong warping, indicating that the Bloch states have a strong symmetry character and enabling the onset of the "3m" torque as well as other unconventional torques. Since the indicator to the presence of this torque is the trigonal warping of the Fermi surface, many other materials could display such effects: For example, Bi-based topological insulators (Bi,Sb)$_2$/(Se,Te)$_{3}$ \cite{Chen2009,Hsieh2009,Alpichshev2010}, and Bi$_4$Te$_3$ \cite{Chagas2022}, but also possibly in the recently grown LaAlO$_3$/EuTiO$_3$/SrTiO$_3$ all-oxide heterostructure \cite{Chen2022}.

We conclude this work by emphasizing that other unconventional torques are yet to be found in low-symmetry crystals that could lead to original current-driven dynamics, as already reported in WTe$_2$/Py \cite{MacNeill2017,Shi2019,Xie2021} and Fe$_3$GeTe$_2$ \cite{Johansen2019,Zhang2021}. In this context, one needs to keep in mind that the general form of the spin-orbit field used in this work is obtained via a low-order expansion of the character table that is formally valid only when the spin-orbit coupling is smaller than the exchange. In materials where the spin-orbit coupling and the exchange are of the same order of magnitude, much more complex torques are expected.

\begin{acknowledgments}
The authors thank L. Liu, J. Chen, K. Belashchenko, J. Medina, J.H. Garcia and S. Roche for fruitful discussions. This work was supported by the ANR ORION project, grant ANR-20-CE30-0022-01 of the French Agence Nationale de la Recherche. In addition, this research was partially financed by the ANR MNEMOSYN project, grant ANR-21-GRF1-0005. D. G.O. and A. M. acknowledge support from the Excellence Initiative of Aix-Marseille Universit\'e - A*Midex, a French "Investissements d'Avenir" program.
\end{acknowledgments}

\bibliography{Biblio2022}

\begin{thebibliography}{66}%
\makeatletter
\providecommand \@ifxundefined [1]{%
 \@ifx{#1\undefined}
}%
\providecommand \@ifnum [1]{%
 \ifnum #1\expandafter \@firstoftwo
 \else \expandafter \@secondoftwo
 \fi
}%
\providecommand \@ifx [1]{%
 \ifx #1\expandafter \@firstoftwo
 \else \expandafter \@secondoftwo
 \fi
}%
\providecommand \natexlab [1]{#1}%
\providecommand \enquote  [1]{``#1''}%
\providecommand \bibnamefont  [1]{#1}%
\providecommand \bibfnamefont [1]{#1}%
\providecommand \citenamefont [1]{#1}%
\providecommand \href@noop [0]{\@secondoftwo}%
\providecommand \href [0]{\begingroup \@sanitize@url \@href}%
\providecommand \@href[1]{\@@startlink{#1}\@@href}%
\providecommand \@@href[1]{\endgroup#1\@@endlink}%
\providecommand \@sanitize@url [0]{\catcode `\\12\catcode `\$12\catcode
  `\&12\catcode `\#12\catcode `\^12\catcode `\_12\catcode `\%12\relax}%
\providecommand \@@startlink[1]{}%
\providecommand \@@endlink[0]{}%
\providecommand \url  [0]{\begingroup\@sanitize@url \@url }%
\providecommand \@url [1]{\endgroup\@href {#1}{\urlprefix }}%
\providecommand \urlprefix  [0]{URL }%
\providecommand \Eprint [0]{\href }%
\providecommand \doibase [0]{https://doi.org/}%
\providecommand \selectlanguage [0]{\@gobble}%
\providecommand \bibinfo  [0]{\@secondoftwo}%
\providecommand \bibfield  [0]{\@secondoftwo}%
\providecommand \translation [1]{[#1]}%
\providecommand \BibitemOpen [0]{}%
\providecommand \bibitemStop [0]{}%
\providecommand \bibitemNoStop [0]{.\EOS\space}%
\providecommand \EOS [0]{\spacefactor3000\relax}%
\providecommand \BibitemShut  [1]{\csname bibitem#1\endcsname}%
\let\auto@bib@innerbib\@empty
\bibitem [{\citenamefont {Manchon}\ \emph {et~al.}(2019)\citenamefont
  {Manchon}, \citenamefont {Zelezn{\'{y}}}, \citenamefont {Miron},
  \citenamefont {Jungwirth}, \citenamefont {Sinova}, \citenamefont {Thiaville},
  \citenamefont {Garello},\ and\ \citenamefont {Gambardella}}]{Manchon2019}%
  \BibitemOpen
  \bibfield  {author} {\bibinfo {author} {\bibfnamefont {A.}~\bibnamefont
  {Manchon}}, \bibinfo {author} {\bibfnamefont {J.}~\bibnamefont
  {Zelezn{\'{y}}}}, \bibinfo {author} {\bibfnamefont {M.}~\bibnamefont
  {Miron}}, \bibinfo {author} {\bibfnamefont {T.}~\bibnamefont {Jungwirth}},
  \bibinfo {author} {\bibfnamefont {J.}~\bibnamefont {Sinova}}, \bibinfo
  {author} {\bibfnamefont {A.}~\bibnamefont {Thiaville}}, \bibinfo {author}
  {\bibfnamefont {K.}~\bibnamefont {Garello}},\ and\ \bibinfo {author}
  {\bibfnamefont {P.}~\bibnamefont {Gambardella}},\ }\bibfield  {title}
  {\bibinfo {title} {{Current-induced spin-orbit torques in ferromagnetic and
  antiferromagnetic systems}},\ }\href
  {https://doi.org/10.1103/RevModPhys.91.035004} {\bibfield  {journal}
  {\bibinfo  {journal} {Review of Modern Physics}\ }\textbf {\bibinfo {volume}
  {91}},\ \bibinfo {pages} {035004} (\bibinfo {year} {2019})}\BibitemShut
  {NoStop}%
\bibitem [{\citenamefont {Miron}\ \emph {et~al.}(2010)\citenamefont {Miron},
  \citenamefont {Gaudin}, \citenamefont {Auffret}, \citenamefont {Rodmacq},
  \citenamefont {Schuhl}, \citenamefont {Pizzini}, \citenamefont {Vogel},\ and\
  \citenamefont {Gambardella}}]{Miron2010}%
  \BibitemOpen
  \bibfield  {author} {\bibinfo {author} {\bibfnamefont {I.~M.}\ \bibnamefont
  {Miron}}, \bibinfo {author} {\bibfnamefont {G.}~\bibnamefont {Gaudin}},
  \bibinfo {author} {\bibfnamefont {S.}~\bibnamefont {Auffret}}, \bibinfo
  {author} {\bibfnamefont {B.}~\bibnamefont {Rodmacq}}, \bibinfo {author}
  {\bibfnamefont {A.}~\bibnamefont {Schuhl}}, \bibinfo {author} {\bibfnamefont
  {S.}~\bibnamefont {Pizzini}}, \bibinfo {author} {\bibfnamefont
  {J.}~\bibnamefont {Vogel}},\ and\ \bibinfo {author} {\bibfnamefont
  {P.}~\bibnamefont {Gambardella}},\ }\bibfield  {title} {\bibinfo {title}
  {{Current-driven spin torque induced by the Rashba effect in a ferromagnetic
  metal layer.}},\ }\href {https://doi.org/10.1038/nmat2613} {\bibfield
  {journal} {\bibinfo  {journal} {Nature Materials}\ }\textbf {\bibinfo
  {volume} {9}},\ \bibinfo {pages} {230} (\bibinfo {year} {2010})}\BibitemShut
  {NoStop}%
\bibitem [{\citenamefont {Liu}\ \emph {et~al.}(2011)\citenamefont {Liu},
  \citenamefont {Moriyama}, \citenamefont {Ralph},\ and\ \citenamefont
  {Buhrman}}]{Liu2011}%
  \BibitemOpen
  \bibfield  {author} {\bibinfo {author} {\bibfnamefont {L.}~\bibnamefont
  {Liu}}, \bibinfo {author} {\bibfnamefont {T.}~\bibnamefont {Moriyama}},
  \bibinfo {author} {\bibfnamefont {D.~C.}\ \bibnamefont {Ralph}},\ and\
  \bibinfo {author} {\bibfnamefont {R.~A.}\ \bibnamefont {Buhrman}},\
  }\bibfield  {title} {\bibinfo {title} {{Spin-Torque Ferromagnetic Resonance
  Induced by the Spin Hall Effect}},\ }\href
  {https://doi.org/10.1103/PhysRevLett.106.036601} {\bibfield  {journal}
  {\bibinfo  {journal} {Physical Review Letters}\ }\textbf {\bibinfo {volume}
  {106}},\ \bibinfo {pages} {036601} (\bibinfo {year} {2011})}\BibitemShut
  {NoStop}%
\bibitem [{\citenamefont {Oh}\ \emph {et~al.}(2016)\citenamefont {Oh},
  \citenamefont {{Chris Baek}}, \citenamefont {Kim}, \citenamefont {Lee},
  \citenamefont {Lee}, \citenamefont {Yang}, \citenamefont {Park},
  \citenamefont {Lee}, \citenamefont {Kim}, \citenamefont {Go}, \citenamefont
  {Jeong}, \citenamefont {Min}, \citenamefont {Lee}, \citenamefont {Lee},\ and\
  \citenamefont {Park}}]{Oh2016}%
  \BibitemOpen
  \bibfield  {author} {\bibinfo {author} {\bibfnamefont {Y.-W.}\ \bibnamefont
  {Oh}}, \bibinfo {author} {\bibfnamefont {S.-H.}\ \bibnamefont {{Chris
  Baek}}}, \bibinfo {author} {\bibfnamefont {Y.~M.}\ \bibnamefont {Kim}},
  \bibinfo {author} {\bibfnamefont {H.~Y.}\ \bibnamefont {Lee}}, \bibinfo
  {author} {\bibfnamefont {K.-D.}\ \bibnamefont {Lee}}, \bibinfo {author}
  {\bibfnamefont {C.-G.}\ \bibnamefont {Yang}}, \bibinfo {author}
  {\bibfnamefont {E.-S.}\ \bibnamefont {Park}}, \bibinfo {author}
  {\bibfnamefont {K.-S.}\ \bibnamefont {Lee}}, \bibinfo {author} {\bibfnamefont
  {K.-W.}\ \bibnamefont {Kim}}, \bibinfo {author} {\bibfnamefont
  {G.}~\bibnamefont {Go}}, \bibinfo {author} {\bibfnamefont {J.-R.}\
  \bibnamefont {Jeong}}, \bibinfo {author} {\bibfnamefont {B.-C.}\ \bibnamefont
  {Min}}, \bibinfo {author} {\bibfnamefont {H.-W.}\ \bibnamefont {Lee}},
  \bibinfo {author} {\bibfnamefont {K.-J.}\ \bibnamefont {Lee}},\ and\ \bibinfo
  {author} {\bibfnamefont {B.-G.}\ \bibnamefont {Park}},\ }\bibfield  {title}
  {\bibinfo {title} {{Field-free switching of perpendicular magnetization
  through spin-orbit torque in antiferromagnet/ferromagnet/oxide
  structures.}},\ }\href {https://doi.org/10.1038/nnano.2016.109} {\bibfield
  {journal} {\bibinfo  {journal} {Nature Nanotechnology}\ }\textbf {\bibinfo
  {volume} {11}},\ \bibinfo {pages} {878} (\bibinfo {year} {2016})}\BibitemShut
  {NoStop}%
\bibitem [{\citenamefont {Fukami}\ \emph {et~al.}(2016)\citenamefont {Fukami},
  \citenamefont {Zhang}, \citenamefont {DuttaGupta},\ and\ \citenamefont
  {Ohno}}]{Fukami2016a}%
  \BibitemOpen
  \bibfield  {author} {\bibinfo {author} {\bibfnamefont {S.}~\bibnamefont
  {Fukami}}, \bibinfo {author} {\bibfnamefont {C.}~\bibnamefont {Zhang}},
  \bibinfo {author} {\bibfnamefont {S.}~\bibnamefont {DuttaGupta}},\ and\
  \bibinfo {author} {\bibfnamefont {H.}~\bibnamefont {Ohno}},\ }\bibfield
  {title} {\bibinfo {title} {{Magnetization switching by spin-orbit torque in
  an antiferromagnet/ferromagnet bilayer system}},\ }\href
  {https://doi.org/10.1038/nmat4566} {\bibfield  {journal} {\bibinfo  {journal}
  {Nature Materials}\ }\textbf {\bibinfo {volume} {15}},\ \bibinfo {pages}
  {535} (\bibinfo {year} {2016})}\BibitemShut {NoStop}%
\bibitem [{\citenamefont {Lau}\ \emph {et~al.}(2016)\citenamefont {Lau},
  \citenamefont {Betto}, \citenamefont {Rode}, \citenamefont {Coey},\ and\
  \citenamefont {Stamenov}}]{Lau2016}%
  \BibitemOpen
  \bibfield  {author} {\bibinfo {author} {\bibfnamefont {Y.-C.}\ \bibnamefont
  {Lau}}, \bibinfo {author} {\bibfnamefont {D.}~\bibnamefont {Betto}}, \bibinfo
  {author} {\bibfnamefont {K.}~\bibnamefont {Rode}}, \bibinfo {author}
  {\bibfnamefont {J.~M.~D.}\ \bibnamefont {Coey}},\ and\ \bibinfo {author}
  {\bibfnamefont {P.}~\bibnamefont {Stamenov}},\ }\bibfield  {title} {\bibinfo
  {title} {{Spin-orbit torque switching without an external field using
  interlayer exchange coupling}},\ }\href
  {https://doi.org/10.1038/nnano.2016.84} {\bibfield  {journal} {\bibinfo
  {journal} {Nature Nanotechnology}\ }\textbf {\bibinfo {volume} {11}},\
  \bibinfo {pages} {758} (\bibinfo {year} {2016})}\BibitemShut {NoStop}%
\bibitem [{\citenamefont {Wei}\ \emph {et~al.}(2022)\citenamefont {Wei},
  \citenamefont {Wang}, \citenamefont {Cui}, \citenamefont {Guo}, \citenamefont
  {Xu}, \citenamefont {Guang}, \citenamefont {Wang}, \citenamefont {Luo},
  \citenamefont {Wan}, \citenamefont {Feng}, \citenamefont {Wei}, \citenamefont
  {Yin}, \citenamefont {Han},\ and\ \citenamefont {Yu}}]{Wei2022}%
  \BibitemOpen
  \bibfield  {author} {\bibinfo {author} {\bibfnamefont {J.}~\bibnamefont
  {Wei}}, \bibinfo {author} {\bibfnamefont {X.}~\bibnamefont {Wang}}, \bibinfo
  {author} {\bibfnamefont {B.}~\bibnamefont {Cui}}, \bibinfo {author}
  {\bibfnamefont {C.}~\bibnamefont {Guo}}, \bibinfo {author} {\bibfnamefont
  {H.}~\bibnamefont {Xu}}, \bibinfo {author} {\bibfnamefont {Y.}~\bibnamefont
  {Guang}}, \bibinfo {author} {\bibfnamefont {Y.}~\bibnamefont {Wang}},
  \bibinfo {author} {\bibfnamefont {X.}~\bibnamefont {Luo}}, \bibinfo {author}
  {\bibfnamefont {C.}~\bibnamefont {Wan}}, \bibinfo {author} {\bibfnamefont
  {J.}~\bibnamefont {Feng}}, \bibinfo {author} {\bibfnamefont {H.}~\bibnamefont
  {Wei}}, \bibinfo {author} {\bibfnamefont {G.}~\bibnamefont {Yin}}, \bibinfo
  {author} {\bibfnamefont {X.}~\bibnamefont {Han}},\ and\ \bibinfo {author}
  {\bibfnamefont {G.}~\bibnamefont {Yu}},\ }\bibfield  {title} {\bibinfo
  {title} {{Field-Free Spin–Orbit Torque Switching in Perpendicularly
  Magnetized Synthetic Antiferromagnets}},\ }\href
  {https://doi.org/10.1002/adfm.202109455} {\bibfield  {journal} {\bibinfo
  {journal} {Advanced Functional Materials}\ }\textbf {\bibinfo {volume}
  {32}},\ \bibinfo {pages} {2109455} (\bibinfo {year} {2022})}\BibitemShut
  {NoStop}%
\bibitem [{\citenamefont {Baek}\ \emph {et~al.}(2018)\citenamefont {Baek},
  \citenamefont {Amin}, \citenamefont {Oh}, \citenamefont {Go}, \citenamefont
  {Lee}, \citenamefont {Lee}, \citenamefont {Kim}, \citenamefont {Stiles},
  \citenamefont {Park},\ and\ \citenamefont {Lee}}]{Baek2018}%
  \BibitemOpen
  \bibfield  {author} {\bibinfo {author} {\bibfnamefont {S.~H.~C.}\
  \bibnamefont {Baek}}, \bibinfo {author} {\bibfnamefont {V.~P.}\ \bibnamefont
  {Amin}}, \bibinfo {author} {\bibfnamefont {Y.~W.}\ \bibnamefont {Oh}},
  \bibinfo {author} {\bibfnamefont {G.}~\bibnamefont {Go}}, \bibinfo {author}
  {\bibfnamefont {S.~J.}\ \bibnamefont {Lee}}, \bibinfo {author} {\bibfnamefont
  {G.~H.}\ \bibnamefont {Lee}}, \bibinfo {author} {\bibfnamefont {K.~J.}\
  \bibnamefont {Kim}}, \bibinfo {author} {\bibfnamefont {M.~D.}\ \bibnamefont
  {Stiles}}, \bibinfo {author} {\bibfnamefont {B.~G.}\ \bibnamefont {Park}},\
  and\ \bibinfo {author} {\bibfnamefont {K.~J.}\ \bibnamefont {Lee}},\
  }\bibfield  {title} {\bibinfo {title} {{Spin currents and spin-orbit torques
  in ferromagnetic trilayers}},\ }\href
  {https://doi.org/10.1038/s41563-018-0041-5} {\bibfield  {journal} {\bibinfo
  {journal} {Nature Materials}\ }\textbf {\bibinfo {volume} {17}},\ \bibinfo
  {pages} {509} (\bibinfo {year} {2018})}\BibitemShut {NoStop}%
\bibitem [{\citenamefont {Ryu}\ \emph {et~al.}(2022)\citenamefont {Ryu},
  \citenamefont {Thompson}, \citenamefont {Park}, \citenamefont {Kim},
  \citenamefont {Choi}, \citenamefont {Kang}, \citenamefont {Jeong},
  \citenamefont {Kohda}, \citenamefont {Yuk}, \citenamefont {Nitta},
  \citenamefont {Lee},\ and\ \citenamefont {Park}}]{Ryu2022}%
  \BibitemOpen
  \bibfield  {author} {\bibinfo {author} {\bibfnamefont {J.}~\bibnamefont
  {Ryu}}, \bibinfo {author} {\bibfnamefont {R.}~\bibnamefont {Thompson}},
  \bibinfo {author} {\bibfnamefont {J.~Y.}\ \bibnamefont {Park}}, \bibinfo
  {author} {\bibfnamefont {S.~J.}\ \bibnamefont {Kim}}, \bibinfo {author}
  {\bibfnamefont {G.}~\bibnamefont {Choi}}, \bibinfo {author} {\bibfnamefont
  {J.}~\bibnamefont {Kang}}, \bibinfo {author} {\bibfnamefont {H.~B.}\
  \bibnamefont {Jeong}}, \bibinfo {author} {\bibfnamefont {M.}~\bibnamefont
  {Kohda}}, \bibinfo {author} {\bibfnamefont {J.~M.}\ \bibnamefont {Yuk}},
  \bibinfo {author} {\bibfnamefont {J.}~\bibnamefont {Nitta}}, \bibinfo
  {author} {\bibfnamefont {K.~J.}\ \bibnamefont {Lee}},\ and\ \bibinfo {author}
  {\bibfnamefont {B.~G.}\ \bibnamefont {Park}},\ }\bibfield  {title} {\bibinfo
  {title} {{Efficient spin–orbit torque in magnetic trilayers using all three
  polarizations of a spin current}},\ }\href
  {https://doi.org/10.1038/s41928-022-00735-9} {\bibfield  {journal} {\bibinfo
  {journal} {Nature Electronics}\ }\textbf {\bibinfo {volume} {5}},\ \bibinfo
  {pages} {217} (\bibinfo {year} {2022})}\BibitemShut {NoStop}%
\bibitem [{\citenamefont {Amin}\ \emph {et~al.}(2018)\citenamefont {Amin},
  \citenamefont {Zemen},\ and\ \citenamefont {Stiles}}]{Amin2018}%
  \BibitemOpen
  \bibfield  {author} {\bibinfo {author} {\bibfnamefont {V.~P.}\ \bibnamefont
  {Amin}}, \bibinfo {author} {\bibfnamefont {J.}~\bibnamefont {Zemen}},\ and\
  \bibinfo {author} {\bibfnamefont {M.~D.}\ \bibnamefont {Stiles}},\ }\bibfield
   {title} {\bibinfo {title} {{Interface-Generated Spin Currents}},\ }\href
  {https://doi.org/10.1103/PhysRevLett.121.136805} {\bibfield  {journal}
  {\bibinfo  {journal} {Physical Review Letters}\ }\textbf {\bibinfo {volume}
  {121}},\ \bibinfo {pages} {136805} (\bibinfo {year} {2018})}\BibitemShut
  {NoStop}%
\bibitem [{\citenamefont {Lifshits}\ and\ \citenamefont
  {Dyakonov}(2009)}]{Lifshits2009}%
  \BibitemOpen
  \bibfield  {author} {\bibinfo {author} {\bibfnamefont {M.~B.}\ \bibnamefont
  {Lifshits}}\ and\ \bibinfo {author} {\bibfnamefont {M.~I.}\ \bibnamefont
  {Dyakonov}},\ }\bibfield  {title} {\bibinfo {title} {{Swapping Spin Currents:
  Interchanging Spin and Flow Directions}},\ }\href
  {https://doi.org/10.1103/PhysRevLett.103.186601} {\bibfield  {journal}
  {\bibinfo  {journal} {Physical Review Letters}\ }\textbf {\bibinfo {volume}
  {103}},\ \bibinfo {pages} {186601} (\bibinfo {year} {2009})}\BibitemShut
  {NoStop}%
\bibitem [{\citenamefont {Saidaoui}\ and\ \citenamefont
  {Manchon}(2016)}]{Saidaoui2016}%
  \BibitemOpen
  \bibfield  {author} {\bibinfo {author} {\bibfnamefont {H.}~\bibnamefont
  {Saidaoui}}\ and\ \bibinfo {author} {\bibfnamefont {A.}~\bibnamefont
  {Manchon}},\ }\bibfield  {title} {\bibinfo {title} {{Spin-Swapping Transport
  and Torques in Ultrathin Magnetic Bilayers}},\ }\href
  {https://doi.org/10.1103/PhysRevLett.117.036601} {\bibfield  {journal}
  {\bibinfo  {journal} {Physical Review Letters}\ }\textbf {\bibinfo {volume}
  {117}},\ \bibinfo {pages} {036601} (\bibinfo {year} {2016})}\BibitemShut
  {NoStop}%
\bibitem [{\citenamefont {Luo}\ \emph {et~al.}(2019)\citenamefont {Luo},
  \citenamefont {Zhang}, \citenamefont {Xu}, \citenamefont {Yang},
  \citenamefont {Zhang},\ and\ \citenamefont {Wu}}]{Luo2019b}%
  \BibitemOpen
  \bibfield  {author} {\bibinfo {author} {\bibfnamefont {Z.}~\bibnamefont
  {Luo}}, \bibinfo {author} {\bibfnamefont {Q.}~\bibnamefont {Zhang}}, \bibinfo
  {author} {\bibfnamefont {Y.}~\bibnamefont {Xu}}, \bibinfo {author}
  {\bibfnamefont {Y.}~\bibnamefont {Yang}}, \bibinfo {author} {\bibfnamefont
  {X.}~\bibnamefont {Zhang}},\ and\ \bibinfo {author} {\bibfnamefont
  {Y.}~\bibnamefont {Wu}},\ }\bibfield  {title} {\bibinfo {title} {{Spin-Orbit
  Torque in a Single Ferromagnetic Layer Induced by Surface Spin Rotation}},\
  }\href {https://doi.org/10.1103/PhysRevApplied.11.064021} {\bibfield
  {journal} {\bibinfo  {journal} {Physical Review Applied}\ }\textbf {\bibinfo
  {volume} {11}},\ \bibinfo {pages} {064021} (\bibinfo {year}
  {2019})}\BibitemShut {NoStop}%
\bibitem [{\citenamefont {Pauyac}\ \emph {et~al.}(2018)\citenamefont {Pauyac},
  \citenamefont {Chshiev}, \citenamefont {Manchon},\ and\ \citenamefont
  {Nikolaev}}]{Pauyac2018}%
  \BibitemOpen
  \bibfield  {author} {\bibinfo {author} {\bibfnamefont {C.~O.}\ \bibnamefont
  {Pauyac}}, \bibinfo {author} {\bibfnamefont {M.}~\bibnamefont {Chshiev}},
  \bibinfo {author} {\bibfnamefont {A.}~\bibnamefont {Manchon}},\ and\ \bibinfo
  {author} {\bibfnamefont {S.~A.}\ \bibnamefont {Nikolaev}},\ }\bibfield
  {title} {\bibinfo {title} {{Spin Hall and Spin Swapping Torques in Diffusive
  Ferromagnets}},\ }\href {https://doi.org/10.1103/PhysRevLett.120.176802}
  {\bibfield  {journal} {\bibinfo  {journal} {Physical Review Letters}\
  }\textbf {\bibinfo {volume} {120}},\ \bibinfo {pages} {176802} (\bibinfo
  {year} {2018})}\BibitemShut {NoStop}%
\bibitem [{\citenamefont {Yu}\ \emph {et~al.}(2014)\citenamefont {Yu},
  \citenamefont {Upadhyaya}, \citenamefont {Fan}, \citenamefont {Alzate},
  \citenamefont {Jiang}, \citenamefont {Wong}, \citenamefont {Takei},
  \citenamefont {Bender}, \citenamefont {Chang}, \citenamefont {Jiang},
  \citenamefont {Lang}, \citenamefont {Tang}, \citenamefont {Wang},
  \citenamefont {Tserkovnyak}, \citenamefont {Amiri},\ and\ \citenamefont
  {Wang}}]{Yu2014}%
  \BibitemOpen
  \bibfield  {author} {\bibinfo {author} {\bibfnamefont {G.}~\bibnamefont
  {Yu}}, \bibinfo {author} {\bibfnamefont {P.}~\bibnamefont {Upadhyaya}},
  \bibinfo {author} {\bibfnamefont {Y.}~\bibnamefont {Fan}}, \bibinfo {author}
  {\bibfnamefont {J.~G.}\ \bibnamefont {Alzate}}, \bibinfo {author}
  {\bibfnamefont {W.}~\bibnamefont {Jiang}}, \bibinfo {author} {\bibfnamefont
  {K.~L.}\ \bibnamefont {Wong}}, \bibinfo {author} {\bibfnamefont
  {S.}~\bibnamefont {Takei}}, \bibinfo {author} {\bibfnamefont {S.~A.}\
  \bibnamefont {Bender}}, \bibinfo {author} {\bibfnamefont {L.-T.}\
  \bibnamefont {Chang}}, \bibinfo {author} {\bibfnamefont {Y.}~\bibnamefont
  {Jiang}}, \bibinfo {author} {\bibfnamefont {M.}~\bibnamefont {Lang}},
  \bibinfo {author} {\bibfnamefont {J.}~\bibnamefont {Tang}}, \bibinfo {author}
  {\bibfnamefont {Y.}~\bibnamefont {Wang}}, \bibinfo {author} {\bibfnamefont
  {Y.}~\bibnamefont {Tserkovnyak}}, \bibinfo {author} {\bibfnamefont {P.~K.}\
  \bibnamefont {Amiri}},\ and\ \bibinfo {author} {\bibfnamefont {K.~L.}\
  \bibnamefont {Wang}},\ }\bibfield  {title} {\bibinfo {title} {{Switching of
  perpendicular magnetization by spin-orbit torques in the absence of external
  magnetic fields.}},\ }\href {https://doi.org/10.1038/nnano.2014.94}
  {\bibfield  {journal} {\bibinfo  {journal} {Nature Nanotechnology}\ }\textbf
  {\bibinfo {volume} {9}},\ \bibinfo {pages} {548} (\bibinfo {year}
  {2014})}\BibitemShut {NoStop}%
\bibitem [{\citenamefont {Bose}\ \emph {et~al.}(2018)\citenamefont {Bose},
  \citenamefont {Lam}, \citenamefont {Bhuktare}, \citenamefont {Dutta},
  \citenamefont {Singh}, \citenamefont {Jibiki},\ and\ \citenamefont
  {Goto}}]{Bose2018a}%
  \BibitemOpen
  \bibfield  {author} {\bibinfo {author} {\bibfnamefont {A.}~\bibnamefont
  {Bose}}, \bibinfo {author} {\bibfnamefont {D.~D.}\ \bibnamefont {Lam}},
  \bibinfo {author} {\bibfnamefont {S.}~\bibnamefont {Bhuktare}}, \bibinfo
  {author} {\bibfnamefont {S.}~\bibnamefont {Dutta}}, \bibinfo {author}
  {\bibfnamefont {H.}~\bibnamefont {Singh}}, \bibinfo {author} {\bibfnamefont
  {Y.}~\bibnamefont {Jibiki}},\ and\ \bibinfo {author} {\bibfnamefont
  {M.}~\bibnamefont {Goto}},\ }\bibfield  {title} {\bibinfo {title}
  {{Observation of Anomalous Spin Torque Generated by a Ferromagnet}},\ }\href
  {https://doi.org/10.1103/PhysRevApplied.9.064026} {\bibfield  {journal}
  {\bibinfo  {journal} {Physical Review Applied}\ }\textbf {\bibinfo {volume}
  {9}},\ \bibinfo {pages} {64026} (\bibinfo {year} {2018})}\BibitemShut
  {NoStop}%
\bibitem [{\citenamefont {Cui}\ \emph {et~al.}(2019)\citenamefont {Cui},
  \citenamefont {Wu}, \citenamefont {Li}, \citenamefont {Razavi}, \citenamefont
  {Wu}, \citenamefont {Wong}, \citenamefont {Chang}, \citenamefont {Gao},
  \citenamefont {Zuo}, \citenamefont {Xi},\ and\ \citenamefont
  {Wang}}]{Cui2019}%
  \BibitemOpen
  \bibfield  {author} {\bibinfo {author} {\bibfnamefont {B.}~\bibnamefont
  {Cui}}, \bibinfo {author} {\bibfnamefont {H.}~\bibnamefont {Wu}}, \bibinfo
  {author} {\bibfnamefont {D.}~\bibnamefont {Li}}, \bibinfo {author}
  {\bibfnamefont {S.~A.}\ \bibnamefont {Razavi}}, \bibinfo {author}
  {\bibfnamefont {D.}~\bibnamefont {Wu}}, \bibinfo {author} {\bibfnamefont
  {K.~L.}\ \bibnamefont {Wong}}, \bibinfo {author} {\bibfnamefont
  {M.}~\bibnamefont {Chang}}, \bibinfo {author} {\bibfnamefont
  {M.}~\bibnamefont {Gao}}, \bibinfo {author} {\bibfnamefont {Y.}~\bibnamefont
  {Zuo}}, \bibinfo {author} {\bibfnamefont {L.}~\bibnamefont {Xi}},\ and\
  \bibinfo {author} {\bibfnamefont {K.~L.}\ \bibnamefont {Wang}},\ }\bibfield
  {title} {\bibinfo {title} {{Field-Free Spin-Orbit Torque Switching of
  Perpendicular Magnetization by the Rashba Interface}},\ }\href
  {https://doi.org/10.1021/acsami.9b13622} {\bibfield  {journal} {\bibinfo
  {journal} {ACS Applied Materials and Interfaces}\ }\textbf {\bibinfo {volume}
  {11}},\ \bibinfo {pages} {39369} (\bibinfo {year} {2019})}\BibitemShut
  {NoStop}%
\bibitem [{\citenamefont {Razavi}\ \emph {et~al.}(2020)\citenamefont {Razavi},
  \citenamefont {Wu}, \citenamefont {Shao}, \citenamefont {Fang}, \citenamefont
  {Dai}, \citenamefont {Wong}, \citenamefont {Han}, \citenamefont {Yu},\ and\
  \citenamefont {Wang}}]{Razari2020}%
  \BibitemOpen
  \bibfield  {author} {\bibinfo {author} {\bibfnamefont {A.}~\bibnamefont
  {Razavi}}, \bibinfo {author} {\bibfnamefont {H.}~\bibnamefont {Wu}}, \bibinfo
  {author} {\bibfnamefont {Q.}~\bibnamefont {Shao}}, \bibinfo {author}
  {\bibfnamefont {C.}~\bibnamefont {Fang}}, \bibinfo {author} {\bibfnamefont
  {B.}~\bibnamefont {Dai}}, \bibinfo {author} {\bibfnamefont {K.}~\bibnamefont
  {Wong}}, \bibinfo {author} {\bibfnamefont {X.}~\bibnamefont {Han}}, \bibinfo
  {author} {\bibfnamefont {G.}~\bibnamefont {Yu}},\ and\ \bibinfo {author}
  {\bibfnamefont {K.~L.}\ \bibnamefont {Wang}},\ }\bibfield  {title} {\bibinfo
  {title} {{Deterministic spin-orbit torque switching by a light-metal
  insertion}},\ }\href {https://doi.org/10.1021/acs.nanolett.0c00647}
  {\bibfield  {journal} {\bibinfo  {journal} {Nano Letters}\ }\textbf {\bibinfo
  {volume} {20}},\ \bibinfo {pages} {3703} (\bibinfo {year}
  {2020})}\BibitemShut {NoStop}%
\bibitem [{\citenamefont {Kang}\ \emph {et~al.}(2021)\citenamefont {Kang},
  \citenamefont {Choi}, \citenamefont {Jeong}, \citenamefont {Park},
  \citenamefont {Park}, \citenamefont {Kim}, \citenamefont {Lee}, \citenamefont
  {Kim}, \citenamefont {Kim}, \citenamefont {Oh}, \citenamefont {Viet},
  \citenamefont {Jeong}, \citenamefont {Yuk}, \citenamefont {Park},
  \citenamefont {Lee},\ and\ \citenamefont {Park}}]{Kang2021}%
  \BibitemOpen
  \bibfield  {author} {\bibinfo {author} {\bibfnamefont {M.~G.}\ \bibnamefont
  {Kang}}, \bibinfo {author} {\bibfnamefont {J.~G.}\ \bibnamefont {Choi}},
  \bibinfo {author} {\bibfnamefont {J.}~\bibnamefont {Jeong}}, \bibinfo
  {author} {\bibfnamefont {J.~Y.}\ \bibnamefont {Park}}, \bibinfo {author}
  {\bibfnamefont {H.~J.}\ \bibnamefont {Park}}, \bibinfo {author}
  {\bibfnamefont {T.}~\bibnamefont {Kim}}, \bibinfo {author} {\bibfnamefont
  {T.}~\bibnamefont {Lee}}, \bibinfo {author} {\bibfnamefont {K.~J.}\
  \bibnamefont {Kim}}, \bibinfo {author} {\bibfnamefont {K.~W.}\ \bibnamefont
  {Kim}}, \bibinfo {author} {\bibfnamefont {J.~H.}\ \bibnamefont {Oh}},
  \bibinfo {author} {\bibfnamefont {D.~D.}\ \bibnamefont {Viet}}, \bibinfo
  {author} {\bibfnamefont {J.~R.}\ \bibnamefont {Jeong}}, \bibinfo {author}
  {\bibfnamefont {J.~M.}\ \bibnamefont {Yuk}}, \bibinfo {author} {\bibfnamefont
  {J.}~\bibnamefont {Park}}, \bibinfo {author} {\bibfnamefont {K.~J.}\
  \bibnamefont {Lee}},\ and\ \bibinfo {author} {\bibfnamefont {B.~G.}\
  \bibnamefont {Park}},\ }\bibfield  {title} {\bibinfo {title} {{Electric-field
  control of field-free spin-orbit torque switching via laterally modulated
  Rashba effect in Pt/Co/AlOx structures}},\ }\href
  {https://doi.org/10.1038/s41467-021-27459-2} {\bibfield  {journal} {\bibinfo
  {journal} {Nature Communications}\ }\textbf {\bibinfo {volume} {12}},\
  \bibinfo {pages} {7111} (\bibinfo {year} {2021})}\BibitemShut {NoStop}%
\bibitem [{\citenamefont {Safeer}\ \emph {et~al.}(2016)\citenamefont {Safeer},
  \citenamefont {Ju{\'{e}}}, \citenamefont {Lopez}, \citenamefont
  {Buda-Prejbeanu}, \citenamefont {Auffret}, \citenamefont {Pizzini},
  \citenamefont {Boulle}, \citenamefont {Miron},\ and\ \citenamefont
  {Gaudin}}]{Safeer2016}%
  \BibitemOpen
  \bibfield  {author} {\bibinfo {author} {\bibfnamefont {C.~K.}\ \bibnamefont
  {Safeer}}, \bibinfo {author} {\bibfnamefont {E.}~\bibnamefont {Ju{\'{e}}}},
  \bibinfo {author} {\bibfnamefont {A.}~\bibnamefont {Lopez}}, \bibinfo
  {author} {\bibfnamefont {L.}~\bibnamefont {Buda-Prejbeanu}}, \bibinfo
  {author} {\bibfnamefont {S.}~\bibnamefont {Auffret}}, \bibinfo {author}
  {\bibfnamefont {S.}~\bibnamefont {Pizzini}}, \bibinfo {author} {\bibfnamefont
  {O.}~\bibnamefont {Boulle}}, \bibinfo {author} {\bibfnamefont {I.~M.}\
  \bibnamefont {Miron}},\ and\ \bibinfo {author} {\bibfnamefont
  {G.}~\bibnamefont {Gaudin}},\ }\bibfield  {title} {\bibinfo {title}
  {{Spin-orbit torque magnetization switching controlled by geometry.}},\
  }\href {https://doi.org/10.1038/nnano.2015.252} {\bibfield  {journal}
  {\bibinfo  {journal} {Nature Nanotechnology}\ }\textbf {\bibinfo {volume}
  {11}},\ \bibinfo {pages} {143} (\bibinfo {year} {2016})}\BibitemShut
  {NoStop}%
\bibitem [{\citenamefont {You}\ \emph {et~al.}(2015)\citenamefont {You},
  \citenamefont {Lee}, \citenamefont {Bhowmik}, \citenamefont {Labanowski},
  \citenamefont {Hong}, \citenamefont {Bokor},\ and\ \citenamefont
  {Salahuddin}}]{You2015}%
  \BibitemOpen
  \bibfield  {author} {\bibinfo {author} {\bibfnamefont {L.}~\bibnamefont
  {You}}, \bibinfo {author} {\bibfnamefont {O.}~\bibnamefont {Lee}}, \bibinfo
  {author} {\bibfnamefont {D.}~\bibnamefont {Bhowmik}}, \bibinfo {author}
  {\bibfnamefont {D.}~\bibnamefont {Labanowski}}, \bibinfo {author}
  {\bibfnamefont {J.}~\bibnamefont {Hong}}, \bibinfo {author} {\bibfnamefont
  {J.}~\bibnamefont {Bokor}},\ and\ \bibinfo {author} {\bibfnamefont
  {S.}~\bibnamefont {Salahuddin}},\ }\bibfield  {title} {\bibinfo {title}
  {{Switching of perpendicularly polarized nanomagnets with spin orbit torque
  without an external magnetic field by engineering a tilted anisotropy}},\
  }\href {https://doi.org/10.1073/pnas.1507474112} {\bibfield  {journal}
  {\bibinfo  {journal} {Proceedings of the National Academy of Sciences}\
  }\textbf {\bibinfo {volume} {112}},\ \bibinfo {pages} {10310} (\bibinfo
  {year} {2015})}\BibitemShut {NoStop}%
\bibitem [{\citenamefont {Liu}\ \emph {et~al.}(2019)\citenamefont {Liu},
  \citenamefont {Qin}, \citenamefont {Lin}, \citenamefont {Li}, \citenamefont
  {Xie}, \citenamefont {He}, \citenamefont {Shu}, \citenamefont {Zhou},
  \citenamefont {Lim}, \citenamefont {Yu}, \citenamefont {Lu}, \citenamefont
  {Li}, \citenamefont {Yan}, \citenamefont {Pennycook},\ and\ \citenamefont
  {Chen}}]{Liu2019f}%
  \BibitemOpen
  \bibfield  {author} {\bibinfo {author} {\bibfnamefont {L.}~\bibnamefont
  {Liu}}, \bibinfo {author} {\bibfnamefont {Q.}~\bibnamefont {Qin}}, \bibinfo
  {author} {\bibfnamefont {W.}~\bibnamefont {Lin}}, \bibinfo {author}
  {\bibfnamefont {C.}~\bibnamefont {Li}}, \bibinfo {author} {\bibfnamefont
  {Q.}~\bibnamefont {Xie}}, \bibinfo {author} {\bibfnamefont {S.}~\bibnamefont
  {He}}, \bibinfo {author} {\bibfnamefont {X.}~\bibnamefont {Shu}}, \bibinfo
  {author} {\bibfnamefont {C.}~\bibnamefont {Zhou}}, \bibinfo {author}
  {\bibfnamefont {Z.}~\bibnamefont {Lim}}, \bibinfo {author} {\bibfnamefont
  {J.}~\bibnamefont {Yu}}, \bibinfo {author} {\bibfnamefont {W.}~\bibnamefont
  {Lu}}, \bibinfo {author} {\bibfnamefont {M.}~\bibnamefont {Li}}, \bibinfo
  {author} {\bibfnamefont {X.}~\bibnamefont {Yan}}, \bibinfo {author}
  {\bibfnamefont {S.~J.}\ \bibnamefont {Pennycook}},\ and\ \bibinfo {author}
  {\bibfnamefont {J.}~\bibnamefont {Chen}},\ }\bibfield  {title} {\bibinfo
  {title} {{Current-induced magnetization switching in all-oxide
  heterostructures}},\ }\href {https://doi.org/10.1038/s41565-019-0534-7}
  {\bibfield  {journal} {\bibinfo  {journal} {Nature Nanotechnology}\ }\textbf
  {\bibinfo {volume} {14}},\ \bibinfo {pages} {939} (\bibinfo {year}
  {2019})}\BibitemShut {NoStop}%
\bibitem [{\citenamefont {Li}\ \emph {et~al.}(2019{\natexlab{a}})\citenamefont
  {Li}, \citenamefont {Wang}, \citenamefont {Li}, \citenamefont {Hu},
  \citenamefont {Zhou}, \citenamefont {Dang}, \citenamefont {Ma}, \citenamefont
  {Dai}, \citenamefont {Kang}, \citenamefont {Yu}, \citenamefont {Zhou},
  \citenamefont {Wu},\ and\ \citenamefont {Li}}]{Li2019l}%
  \BibitemOpen
  \bibfield  {author} {\bibinfo {author} {\bibfnamefont {H.}~\bibnamefont
  {Li}}, \bibinfo {author} {\bibfnamefont {G.}~\bibnamefont {Wang}}, \bibinfo
  {author} {\bibfnamefont {D.}~\bibnamefont {Li}}, \bibinfo {author}
  {\bibfnamefont {P.}~\bibnamefont {Hu}}, \bibinfo {author} {\bibfnamefont
  {W.}~\bibnamefont {Zhou}}, \bibinfo {author} {\bibfnamefont {S.}~\bibnamefont
  {Dang}}, \bibinfo {author} {\bibfnamefont {X.}~\bibnamefont {Ma}}, \bibinfo
  {author} {\bibfnamefont {T.}~\bibnamefont {Dai}}, \bibinfo {author}
  {\bibfnamefont {S.}~\bibnamefont {Kang}}, \bibinfo {author} {\bibfnamefont
  {F.}~\bibnamefont {Yu}}, \bibinfo {author} {\bibfnamefont {X.}~\bibnamefont
  {Zhou}}, \bibinfo {author} {\bibfnamefont {S.}~\bibnamefont {Wu}},\ and\
  \bibinfo {author} {\bibfnamefont {S.}~\bibnamefont {Li}},\ }\bibfield
  {title} {\bibinfo {title} {{Field-Free Deterministic Magnetization Switching
  with Ultralow Current Density in Epitaxial Au/Fe 4 N Bilayer Films}},\ }\href
  {https://doi.org/10.1021/acsami.9b00129} {\bibfield  {journal} {\bibinfo
  {journal} {ACS Applied Materials and Interfaces}\ }\textbf {\bibinfo {volume}
  {11}},\ \bibinfo {pages} {16965} (\bibinfo {year}
  {2019}{\natexlab{a}})}\BibitemShut {NoStop}%
\bibitem [{\citenamefont {Shu}\ \emph {et~al.}(2022)\citenamefont {Shu},
  \citenamefont {Liu}, \citenamefont {Zhou}, \citenamefont {Lin}, \citenamefont
  {Xie}, \citenamefont {Zhao}, \citenamefont {Zhou}, \citenamefont {Chen},
  \citenamefont {Wang}, \citenamefont {Chai}, \citenamefont {Ding},
  \citenamefont {Chen},\ and\ \citenamefont {Chen}}]{Shu2022}%
  \BibitemOpen
  \bibfield  {author} {\bibinfo {author} {\bibfnamefont {X.}~\bibnamefont
  {Shu}}, \bibinfo {author} {\bibfnamefont {L.}~\bibnamefont {Liu}}, \bibinfo
  {author} {\bibfnamefont {J.}~\bibnamefont {Zhou}}, \bibinfo {author}
  {\bibfnamefont {W.}~\bibnamefont {Lin}}, \bibinfo {author} {\bibfnamefont
  {Q.}~\bibnamefont {Xie}}, \bibinfo {author} {\bibfnamefont {T.}~\bibnamefont
  {Zhao}}, \bibinfo {author} {\bibfnamefont {C.}~\bibnamefont {Zhou}}, \bibinfo
  {author} {\bibfnamefont {S.}~\bibnamefont {Chen}}, \bibinfo {author}
  {\bibfnamefont {H.}~\bibnamefont {Wang}}, \bibinfo {author} {\bibfnamefont
  {J.}~\bibnamefont {Chai}}, \bibinfo {author} {\bibfnamefont {Y.}~\bibnamefont
  {Ding}}, \bibinfo {author} {\bibfnamefont {W.}~\bibnamefont {Chen}},\ and\
  \bibinfo {author} {\bibfnamefont {J.}~\bibnamefont {Chen}},\ }\bibfield
  {title} {\bibinfo {title} {{Field-Free Switching of Perpendicular
  Magnetization Induced by Longitudinal Spin-Orbit-Torque Gradient}},\ }\href
  {https://doi.org/10.1103/PhysRevApplied.17.024031} {\bibfield  {journal}
  {\bibinfo  {journal} {Physical Review Applied}\ }\textbf {\bibinfo {volume}
  {17}},\ \bibinfo {pages} {024031} (\bibinfo {year} {2022})}\BibitemShut
  {NoStop}%
\bibitem [{\citenamefont {Chen}\ \emph {et~al.}(2019)\citenamefont {Chen},
  \citenamefont {Yu}, \citenamefont {Xie}, \citenamefont {Zhang}, \citenamefont
  {Lin}, \citenamefont {Liu}, \citenamefont {Zhou}, \citenamefont {Shu},
  \citenamefont {Guo}, \citenamefont {Zhang},\ and\ \citenamefont
  {Chen}}]{Chen2019c}%
  \BibitemOpen
  \bibfield  {author} {\bibinfo {author} {\bibfnamefont {S.}~\bibnamefont
  {Chen}}, \bibinfo {author} {\bibfnamefont {J.}~\bibnamefont {Yu}}, \bibinfo
  {author} {\bibfnamefont {Q.}~\bibnamefont {Xie}}, \bibinfo {author}
  {\bibfnamefont {X.}~\bibnamefont {Zhang}}, \bibinfo {author} {\bibfnamefont
  {W.}~\bibnamefont {Lin}}, \bibinfo {author} {\bibfnamefont {L.}~\bibnamefont
  {Liu}}, \bibinfo {author} {\bibfnamefont {J.}~\bibnamefont {Zhou}}, \bibinfo
  {author} {\bibfnamefont {X.}~\bibnamefont {Shu}}, \bibinfo {author}
  {\bibfnamefont {R.}~\bibnamefont {Guo}}, \bibinfo {author} {\bibfnamefont
  {Z.}~\bibnamefont {Zhang}},\ and\ \bibinfo {author} {\bibfnamefont
  {J.}~\bibnamefont {Chen}},\ }\bibfield  {title} {\bibinfo {title} {{Free
  Field Electric Switching of Perpendicularly Magnetized Thin Film by Spin
  Current Gradient}},\ }\href {https://doi.org/10.1021/acsami.9b09146}
  {\bibfield  {journal} {\bibinfo  {journal} {ACS Applied Materials and
  Interfaces}\ }\textbf {\bibinfo {volume} {11}},\ \bibinfo {pages} {30446}
  (\bibinfo {year} {2019})}\BibitemShut {NoStop}%
\bibitem [{\citenamefont {Chernyshov}\ \emph {et~al.}(2009)\citenamefont
  {Chernyshov}, \citenamefont {Overby}, \citenamefont {Liu}, \citenamefont
  {Furdyna}, \citenamefont {Lyanda-Geller},\ and\ \citenamefont
  {Rokhinson}}]{Chernyshov2009}%
  \BibitemOpen
  \bibfield  {author} {\bibinfo {author} {\bibfnamefont {A.}~\bibnamefont
  {Chernyshov}}, \bibinfo {author} {\bibfnamefont {M.}~\bibnamefont {Overby}},
  \bibinfo {author} {\bibfnamefont {X.}~\bibnamefont {Liu}}, \bibinfo {author}
  {\bibfnamefont {J.~K.}\ \bibnamefont {Furdyna}}, \bibinfo {author}
  {\bibfnamefont {Y.}~\bibnamefont {Lyanda-Geller}},\ and\ \bibinfo {author}
  {\bibfnamefont {L.~P.}\ \bibnamefont {Rokhinson}},\ }\bibfield  {title}
  {\bibinfo {title} {{Evidence for reversible control of magnetization in a
  ferromagnetic material by means of spin–orbit magnetic field}},\ }\href
  {https://doi.org/10.1038/nphys1362} {\bibfield  {journal} {\bibinfo
  {journal} {Nature Physics}\ }\textbf {\bibinfo {volume} {5}},\ \bibinfo
  {pages} {656} (\bibinfo {year} {2009})}\BibitemShut {NoStop}%
\bibitem [{\citenamefont {Endo}\ \emph {et~al.}(2010)\citenamefont {Endo},
  \citenamefont {Matsukura},\ and\ \citenamefont {Ohno}}]{Endo2010}%
  \BibitemOpen
  \bibfield  {author} {\bibinfo {author} {\bibfnamefont {M.}~\bibnamefont
  {Endo}}, \bibinfo {author} {\bibfnamefont {F.}~\bibnamefont {Matsukura}},\
  and\ \bibinfo {author} {\bibfnamefont {H.}~\bibnamefont {Ohno}},\ }\bibfield
  {title} {\bibinfo {title} {{Current induced effective magnetic field and
  magnetization reversal in uniaxial anisotropy (Ga,Mn)As}},\ }\href
  {https://doi.org/10.1063/1.3520514} {\bibfield  {journal} {\bibinfo
  {journal} {Applied Physics Letters}\ }\textbf {\bibinfo {volume} {97}},\
  \bibinfo {pages} {222501} (\bibinfo {year} {2010})}\BibitemShut {NoStop}%
\bibitem [{\citenamefont {Ciccarelli}\ \emph {et~al.}(2016)\citenamefont
  {Ciccarelli}, \citenamefont {Anderson}, \citenamefont {Tshitoyan},
  \citenamefont {Ferguson}, \citenamefont {Gerhard}, \citenamefont {Gould},
  \citenamefont {Molenkamp}, \citenamefont {Gayles}, \citenamefont {Zelezny},
  \citenamefont {Smejkal}, \citenamefont {Yuan}, \citenamefont {Sinova},
  \citenamefont {Freimuth},\ and\ \citenamefont {Jungwirth}}]{Ciccarelli2016}%
  \BibitemOpen
  \bibfield  {author} {\bibinfo {author} {\bibfnamefont {C.}~\bibnamefont
  {Ciccarelli}}, \bibinfo {author} {\bibfnamefont {L.}~\bibnamefont
  {Anderson}}, \bibinfo {author} {\bibfnamefont {V.}~\bibnamefont {Tshitoyan}},
  \bibinfo {author} {\bibfnamefont {A.~J.}\ \bibnamefont {Ferguson}}, \bibinfo
  {author} {\bibfnamefont {F.}~\bibnamefont {Gerhard}}, \bibinfo {author}
  {\bibfnamefont {C.}~\bibnamefont {Gould}}, \bibinfo {author} {\bibfnamefont
  {L.~W.}\ \bibnamefont {Molenkamp}}, \bibinfo {author} {\bibfnamefont
  {J.}~\bibnamefont {Gayles}}, \bibinfo {author} {\bibfnamefont
  {J.}~\bibnamefont {Zelezny}}, \bibinfo {author} {\bibfnamefont
  {L.}~\bibnamefont {Smejkal}}, \bibinfo {author} {\bibfnamefont
  {Z.}~\bibnamefont {Yuan}}, \bibinfo {author} {\bibfnamefont {J.}~\bibnamefont
  {Sinova}}, \bibinfo {author} {\bibfnamefont {F.}~\bibnamefont {Freimuth}},\
  and\ \bibinfo {author} {\bibfnamefont {T.}~\bibnamefont {Jungwirth}},\
  }\bibfield  {title} {\bibinfo {title} {{Room-temperature spin-orbit torque in
  NiMnSb}},\ }\href {https://doi.org/10.1038/nphys3772} {\bibfield  {journal}
  {\bibinfo  {journal} {Nature Physics}\ }\textbf {\bibinfo {volume} {12}},\
  \bibinfo {pages} {855} (\bibinfo {year} {2016})}\BibitemShut {NoStop}%
\bibitem [{\citenamefont {MacNeill}\ \emph {et~al.}(2017)\citenamefont
  {MacNeill}, \citenamefont {Stiehl}, \citenamefont {Guimaraes}, \citenamefont
  {Buhrman}, \citenamefont {Park},\ and\ \citenamefont {Ralph}}]{MacNeill2017}%
  \BibitemOpen
  \bibfield  {author} {\bibinfo {author} {\bibfnamefont {D.}~\bibnamefont
  {MacNeill}}, \bibinfo {author} {\bibfnamefont {G.~M.}\ \bibnamefont
  {Stiehl}}, \bibinfo {author} {\bibfnamefont {M.~H.~D.}\ \bibnamefont
  {Guimaraes}}, \bibinfo {author} {\bibfnamefont {R.~A.}\ \bibnamefont
  {Buhrman}}, \bibinfo {author} {\bibfnamefont {J.}~\bibnamefont {Park}},\ and\
  \bibinfo {author} {\bibfnamefont {D.~C.}\ \bibnamefont {Ralph}},\ }\bibfield
  {title} {\bibinfo {title} {{Control of spin-orbit torques through crystal
  symmetry in WTe2/ferromagnet bilayers}},\ }\href
  {https://doi.org/10.1038/nphys3933} {\bibfield  {journal} {\bibinfo
  {journal} {Nature Physics}\ }\textbf {\bibinfo {volume} {13}},\ \bibinfo
  {pages} {300} (\bibinfo {year} {2017})}\BibitemShut {NoStop}%
\bibitem [{\citenamefont {Macneill}\ \emph {et~al.}(2017)\citenamefont
  {Macneill}, \citenamefont {Stiehl}, \citenamefont {Guimar{\~{a}}es},
  \citenamefont {Reynolds}, \citenamefont {Buhrman},\ and\ \citenamefont
  {Ralph}}]{MacNeill2017b}%
  \BibitemOpen
  \bibfield  {author} {\bibinfo {author} {\bibfnamefont {D.}~\bibnamefont
  {Macneill}}, \bibinfo {author} {\bibfnamefont {G.~M.}\ \bibnamefont
  {Stiehl}}, \bibinfo {author} {\bibfnamefont {M.~H.}\ \bibnamefont
  {Guimar{\~{a}}es}}, \bibinfo {author} {\bibfnamefont {N.~D.}\ \bibnamefont
  {Reynolds}}, \bibinfo {author} {\bibfnamefont {R.~A.}\ \bibnamefont
  {Buhrman}},\ and\ \bibinfo {author} {\bibfnamefont {D.~C.}\ \bibnamefont
  {Ralph}},\ }\bibfield  {title} {\bibinfo {title} {{Thickness dependence of
  spin-orbit torques generated by WTe2}},\ }\href
  {https://doi.org/10.1103/PhysRevB.96.054450} {\bibfield  {journal} {\bibinfo
  {journal} {Physical Review B}\ }\textbf {\bibinfo {volume} {96}},\ \bibinfo
  {pages} {054450} (\bibinfo {year} {2017})},\ \Eprint
  {https://arxiv.org/abs/1707.03757} {arXiv:1707.03757} \BibitemShut {NoStop}%
\bibitem [{\citenamefont {Shi}\ \emph {et~al.}(2019)\citenamefont {Shi},
  \citenamefont {Liang}, \citenamefont {Zhu}, \citenamefont {Cai},
  \citenamefont {Pollard}, \citenamefont {Wang}, \citenamefont {Wang},
  \citenamefont {Wang}, \citenamefont {He}, \citenamefont {Yu}, \citenamefont
  {Eda}, \citenamefont {Liang},\ and\ \citenamefont {Yang}}]{Shi2019}%
  \BibitemOpen
  \bibfield  {author} {\bibinfo {author} {\bibfnamefont {S.}~\bibnamefont
  {Shi}}, \bibinfo {author} {\bibfnamefont {S.}~\bibnamefont {Liang}}, \bibinfo
  {author} {\bibfnamefont {Z.}~\bibnamefont {Zhu}}, \bibinfo {author}
  {\bibfnamefont {K.}~\bibnamefont {Cai}}, \bibinfo {author} {\bibfnamefont
  {S.~D.}\ \bibnamefont {Pollard}}, \bibinfo {author} {\bibfnamefont
  {Y.}~\bibnamefont {Wang}}, \bibinfo {author} {\bibfnamefont {J.}~\bibnamefont
  {Wang}}, \bibinfo {author} {\bibfnamefont {Q.}~\bibnamefont {Wang}}, \bibinfo
  {author} {\bibfnamefont {P.}~\bibnamefont {He}}, \bibinfo {author}
  {\bibfnamefont {J.}~\bibnamefont {Yu}}, \bibinfo {author} {\bibfnamefont
  {G.}~\bibnamefont {Eda}}, \bibinfo {author} {\bibfnamefont {G.}~\bibnamefont
  {Liang}},\ and\ \bibinfo {author} {\bibfnamefont {H.}~\bibnamefont {Yang}},\
  }\bibfield  {title} {\bibinfo {title} {{All-electric magnetization switching
  and Dzyaloshinskii–Moriya interaction in WTe2/ferromagnet
  heterostructures}},\ }\href {https://doi.org/10.1038/s41565-019-0525-8}
  {\bibfield  {journal} {\bibinfo  {journal} {Nature Nanotechnology}\ }\textbf
  {\bibinfo {volume} {14}},\ \bibinfo {pages} {945} (\bibinfo {year}
  {2019})}\BibitemShut {NoStop}%
\bibitem [{\citenamefont {Xie}\ \emph {et~al.}(2021)\citenamefont {Xie},
  \citenamefont {Lin}, \citenamefont {Sarkar}, \citenamefont {Shu},
  \citenamefont {Chen}, \citenamefont {Liu}, \citenamefont {Zhao},
  \citenamefont {Zhou}, \citenamefont {Wang}, \citenamefont {Zhou},
  \citenamefont {Grade{\v{c}}ak},\ and\ \citenamefont {Chen}}]{Xie2021}%
  \BibitemOpen
  \bibfield  {author} {\bibinfo {author} {\bibfnamefont {Q.}~\bibnamefont
  {Xie}}, \bibinfo {author} {\bibfnamefont {W.}~\bibnamefont {Lin}}, \bibinfo
  {author} {\bibfnamefont {S.}~\bibnamefont {Sarkar}}, \bibinfo {author}
  {\bibfnamefont {X.}~\bibnamefont {Shu}}, \bibinfo {author} {\bibfnamefont
  {S.}~\bibnamefont {Chen}}, \bibinfo {author} {\bibfnamefont {L.}~\bibnamefont
  {Liu}}, \bibinfo {author} {\bibfnamefont {T.}~\bibnamefont {Zhao}}, \bibinfo
  {author} {\bibfnamefont {C.}~\bibnamefont {Zhou}}, \bibinfo {author}
  {\bibfnamefont {H.}~\bibnamefont {Wang}}, \bibinfo {author} {\bibfnamefont
  {J.}~\bibnamefont {Zhou}}, \bibinfo {author} {\bibfnamefont {S.}~\bibnamefont
  {Grade{\v{c}}ak}},\ and\ \bibinfo {author} {\bibfnamefont {J.}~\bibnamefont
  {Chen}},\ }\bibfield  {title} {\bibinfo {title} {{Field-free magnetization
  switching induced by the unconventional spin-orbit torque from WTe2}},\
  }\href {https://doi.org/10.1063/5.0048926} {\bibfield  {journal} {\bibinfo
  {journal} {APL Materials}\ }\textbf {\bibinfo {volume} {9}},\ \bibinfo
  {pages} {051114} (\bibinfo {year} {2021})}\BibitemShut {NoStop}%
\bibitem [{\citenamefont {Kao}\ \emph {et~al.}(2022)\citenamefont {Kao},
  \citenamefont {Muzzio}, \citenamefont {Zhang}, \citenamefont {Zhu},
  \citenamefont {Gobbo}, \citenamefont {Yuan}, \citenamefont {Weber},
  \citenamefont {Rao}, \citenamefont {Li}, \citenamefont {Edgar}, \citenamefont
  {Goldberger}, \citenamefont {Yan}, \citenamefont {Mandrus}, \citenamefont
  {Hwang}, \citenamefont {Cheng}, \citenamefont {Katoch},\ and\ \citenamefont
  {Singh}}]{Kao2022}%
  \BibitemOpen
  \bibfield  {author} {\bibinfo {author} {\bibfnamefont {I.-h.}\ \bibnamefont
  {Kao}}, \bibinfo {author} {\bibfnamefont {R.}~\bibnamefont {Muzzio}},
  \bibinfo {author} {\bibfnamefont {H.}~\bibnamefont {Zhang}}, \bibinfo
  {author} {\bibfnamefont {M.}~\bibnamefont {Zhu}}, \bibinfo {author}
  {\bibfnamefont {J.}~\bibnamefont {Gobbo}}, \bibinfo {author} {\bibfnamefont
  {S.}~\bibnamefont {Yuan}}, \bibinfo {author} {\bibfnamefont {D.}~\bibnamefont
  {Weber}}, \bibinfo {author} {\bibfnamefont {R.}~\bibnamefont {Rao}}, \bibinfo
  {author} {\bibfnamefont {J.}~\bibnamefont {Li}}, \bibinfo {author}
  {\bibfnamefont {J.~H.}\ \bibnamefont {Edgar}}, \bibinfo {author}
  {\bibfnamefont {J.~E.}\ \bibnamefont {Goldberger}}, \bibinfo {author}
  {\bibfnamefont {J.}~\bibnamefont {Yan}}, \bibinfo {author} {\bibfnamefont
  {D.~G.}\ \bibnamefont {Mandrus}}, \bibinfo {author} {\bibfnamefont
  {J.}~\bibnamefont {Hwang}}, \bibinfo {author} {\bibfnamefont
  {R.}~\bibnamefont {Cheng}}, \bibinfo {author} {\bibfnamefont
  {J.}~\bibnamefont {Katoch}},\ and\ \bibinfo {author} {\bibfnamefont
  {S.}~\bibnamefont {Singh}},\ }\bibfield  {title} {\bibinfo {title}
  {{Deterministic switching of a perpendicularly polarized magnet using
  unconventional spin – orbit torques in WTe2}},\ }\bibfield  {journal}
  {\bibinfo  {journal} {Nature Materials}\ }\href
  {https://doi.org/10.1038/s41563-022-01275-5} {10.1038/s41563-022-01275-5}
  (\bibinfo {year} {2022})\BibitemShut {NoStop}%
\bibitem [{\citenamefont {Xue}\ \emph {et~al.}(2020)\citenamefont {Xue},
  \citenamefont {Rohmann}, \citenamefont {Li}, \citenamefont {Amin},\ and\
  \citenamefont {Haney}}]{Xue2020}%
  \BibitemOpen
  \bibfield  {author} {\bibinfo {author} {\bibfnamefont {F.}~\bibnamefont
  {Xue}}, \bibinfo {author} {\bibfnamefont {C.}~\bibnamefont {Rohmann}},
  \bibinfo {author} {\bibfnamefont {J.}~\bibnamefont {Li}}, \bibinfo {author}
  {\bibfnamefont {V.}~\bibnamefont {Amin}},\ and\ \bibinfo {author}
  {\bibfnamefont {P.}~\bibnamefont {Haney}},\ }\bibfield  {title} {\bibinfo
  {title} {{Unconventional spin-orbit torque in transition metal
  dichalcogenide-ferromagnet bilayers from first-principles calculations}},\
  }\href {https://doi.org/10.1103/PhysRevB.102.014401} {\bibfield  {journal}
  {\bibinfo  {journal} {Physical Review B}\ }\textbf {\bibinfo {volume}
  {102}},\ \bibinfo {pages} {014401} (\bibinfo {year} {2020})},\ \Eprint
  {https://arxiv.org/abs/2005.01109} {arXiv:2005.01109} \BibitemShut {NoStop}%
\bibitem [{\citenamefont {Guimar{\~{a}}es}\ \emph {et~al.}(2018)\citenamefont
  {Guimar{\~{a}}es}, \citenamefont {Stiehl}, \citenamefont {MacNeill},
  \citenamefont {Reynolds},\ and\ \citenamefont {Ralph}}]{Guimaraes2018}%
  \BibitemOpen
  \bibfield  {author} {\bibinfo {author} {\bibfnamefont {M.~H.}\ \bibnamefont
  {Guimar{\~{a}}es}}, \bibinfo {author} {\bibfnamefont {G.~M.}\ \bibnamefont
  {Stiehl}}, \bibinfo {author} {\bibfnamefont {D.}~\bibnamefont {MacNeill}},
  \bibinfo {author} {\bibfnamefont {N.~D.}\ \bibnamefont {Reynolds}},\ and\
  \bibinfo {author} {\bibfnamefont {D.~C.}\ \bibnamefont {Ralph}},\ }\bibfield
  {title} {\bibinfo {title} {{Spin-Orbit Torques in NbSe2/Permalloy
  Bilayers}},\ }\href {https://doi.org/10.1021/acs.nanolett.7b04993} {\bibfield
   {journal} {\bibinfo  {journal} {Nano Letters}\ }\textbf {\bibinfo {volume}
  {18}},\ \bibinfo {pages} {1311} (\bibinfo {year} {2018})}\BibitemShut
  {NoStop}%
\bibitem [{\citenamefont {Zelezn{\'{y}}}\ \emph {et~al.}(2017)\citenamefont
  {Zelezn{\'{y}}}, \citenamefont {Zhang}, \citenamefont {Felser},\ and\
  \citenamefont {Yan}}]{Zelezny2017b}%
  \BibitemOpen
  \bibfield  {author} {\bibinfo {author} {\bibfnamefont {J.}~\bibnamefont
  {Zelezn{\'{y}}}}, \bibinfo {author} {\bibfnamefont {Y.}~\bibnamefont
  {Zhang}}, \bibinfo {author} {\bibfnamefont {C.}~\bibnamefont {Felser}},\ and\
  \bibinfo {author} {\bibfnamefont {B.}~\bibnamefont {Yan}},\ }\bibfield
  {title} {\bibinfo {title} {{Spin-Polarized Current in Noncollinear
  Antiferromagnets}},\ }\href {https://doi.org/10.1103/PhysRevLett.119.187204}
  {\bibfield  {journal} {\bibinfo  {journal} {Physical Review Letters}\
  }\textbf {\bibinfo {volume} {119}},\ \bibinfo {pages} {187204} (\bibinfo
  {year} {2017})}\BibitemShut {NoStop}%
\bibitem [{\citenamefont {Zhang}\ \emph {et~al.}(2018)\citenamefont {Zhang},
  \citenamefont {Zelezny}, \citenamefont {Sun}, \citenamefont {Brink},\ and\
  \citenamefont {Yan}}]{Zhang2018d}%
  \BibitemOpen
  \bibfield  {author} {\bibinfo {author} {\bibfnamefont {Y.}~\bibnamefont
  {Zhang}}, \bibinfo {author} {\bibfnamefont {J.}~\bibnamefont {Zelezny}},
  \bibinfo {author} {\bibfnamefont {Y.}~\bibnamefont {Sun}}, \bibinfo {author}
  {\bibfnamefont {J.~V.~D.}\ \bibnamefont {Brink}},\ and\ \bibinfo {author}
  {\bibfnamefont {B.}~\bibnamefont {Yan}},\ }\bibfield  {title} {\bibinfo
  {title} {{Spin Hall effect emerging from a noncollinear magnetic lattice
  without spin-orbit coupling}},\ }\href@noop {} {\bibfield  {journal}
  {\bibinfo  {journal} {New Journal of Physics}\ }\textbf {\bibinfo {volume}
  {20}},\ \bibinfo {pages} {073028} (\bibinfo {year} {2018})}\BibitemShut
  {NoStop}%
\bibitem [{\citenamefont {Kimata}\ \emph {et~al.}(2019)\citenamefont {Kimata},
  \citenamefont {Chen}, \citenamefont {Kondou}, \citenamefont {Sugimoto},
  \citenamefont {Muduli}, \citenamefont {Ikhlas}, \citenamefont {Omori},
  \citenamefont {Tomita}, \citenamefont {Macdonald}, \citenamefont
  {Nakatsuji},\ and\ \citenamefont {Otani}}]{Kimata2019}%
  \BibitemOpen
  \bibfield  {author} {\bibinfo {author} {\bibfnamefont {M.}~\bibnamefont
  {Kimata}}, \bibinfo {author} {\bibfnamefont {H.}~\bibnamefont {Chen}},
  \bibinfo {author} {\bibfnamefont {K.}~\bibnamefont {Kondou}}, \bibinfo
  {author} {\bibfnamefont {S.}~\bibnamefont {Sugimoto}}, \bibinfo {author}
  {\bibfnamefont {P.~K.}\ \bibnamefont {Muduli}}, \bibinfo {author}
  {\bibfnamefont {M.}~\bibnamefont {Ikhlas}}, \bibinfo {author} {\bibfnamefont
  {Y.}~\bibnamefont {Omori}}, \bibinfo {author} {\bibfnamefont
  {T.}~\bibnamefont {Tomita}}, \bibinfo {author} {\bibfnamefont {A.~H.}\
  \bibnamefont {Macdonald}}, \bibinfo {author} {\bibfnamefont {S.}~\bibnamefont
  {Nakatsuji}},\ and\ \bibinfo {author} {\bibfnamefont {Y.}~\bibnamefont
  {Otani}},\ }\bibfield  {title} {\bibinfo {title} {{Magnetic and magnetic
  inverse spin Hall effects in a non-collinear antiferromagnet}},\ }\href
  {https://doi.org/10.1038/s41586-018-0853-0} {\bibfield  {journal} {\bibinfo
  {journal} {Nature}\ }\textbf {\bibinfo {volume} {565}},\ \bibinfo {pages}
  {627} (\bibinfo {year} {2019})}\BibitemShut {NoStop}%
\bibitem [{\citenamefont {Ghosh}\ \emph {et~al.}(2022)\citenamefont {Ghosh},
  \citenamefont {Manchon},\ and\ \citenamefont
  {{\v{Z}}elezn{\'{y}}}}]{Ghosh2022}%
  \BibitemOpen
  \bibfield  {author} {\bibinfo {author} {\bibfnamefont {S.}~\bibnamefont
  {Ghosh}}, \bibinfo {author} {\bibfnamefont {A.}~\bibnamefont {Manchon}},\
  and\ \bibinfo {author} {\bibfnamefont {J.}~\bibnamefont
  {{\v{Z}}elezn{\'{y}}}},\ }\bibfield  {title} {\bibinfo {title}
  {{Unconventional Robust Spin-Transfer Torque in Noncollinear
  Antiferromagnetic Junctions}},\ }\href
  {https://doi.org/10.1103/physrevlett.128.097702} {\bibfield  {journal}
  {\bibinfo  {journal} {Physical Review Letters}\ }\textbf {\bibinfo {volume}
  {128}},\ \bibinfo {pages} {097702} (\bibinfo {year} {2022})}\BibitemShut
  {NoStop}%
\bibitem [{\citenamefont {Chen}\ \emph {et~al.}(2021)\citenamefont {Chen},
  \citenamefont {Shi}, \citenamefont {Shi}, \citenamefont {Fan}, \citenamefont
  {Song}, \citenamefont {Zhou}, \citenamefont {Bai}, \citenamefont {Liao},
  \citenamefont {Zhou}, \citenamefont {Zhang}, \citenamefont {Li},
  \citenamefont {Chen}, \citenamefont {Han}, \citenamefont {Jiang},
  \citenamefont {Zhu}, \citenamefont {Wu}, \citenamefont {Wang}, \citenamefont
  {Xue}, \citenamefont {Yang},\ and\ \citenamefont {Pan}}]{Chen2020c}%
  \BibitemOpen
  \bibfield  {author} {\bibinfo {author} {\bibfnamefont {X.}~\bibnamefont
  {Chen}}, \bibinfo {author} {\bibfnamefont {S.}~\bibnamefont {Shi}}, \bibinfo
  {author} {\bibfnamefont {G.}~\bibnamefont {Shi}}, \bibinfo {author}
  {\bibfnamefont {X.}~\bibnamefont {Fan}}, \bibinfo {author} {\bibfnamefont
  {C.}~\bibnamefont {Song}}, \bibinfo {author} {\bibfnamefont {X.}~\bibnamefont
  {Zhou}}, \bibinfo {author} {\bibfnamefont {H.}~\bibnamefont {Bai}}, \bibinfo
  {author} {\bibfnamefont {L.}~\bibnamefont {Liao}}, \bibinfo {author}
  {\bibfnamefont {Y.}~\bibnamefont {Zhou}}, \bibinfo {author} {\bibfnamefont
  {H.}~\bibnamefont {Zhang}}, \bibinfo {author} {\bibfnamefont
  {A.}~\bibnamefont {Li}}, \bibinfo {author} {\bibfnamefont {Y.}~\bibnamefont
  {Chen}}, \bibinfo {author} {\bibfnamefont {X.}~\bibnamefont {Han}}, \bibinfo
  {author} {\bibfnamefont {S.}~\bibnamefont {Jiang}}, \bibinfo {author}
  {\bibfnamefont {Z.}~\bibnamefont {Zhu}}, \bibinfo {author} {\bibfnamefont
  {H.}~\bibnamefont {Wu}}, \bibinfo {author} {\bibfnamefont {X.}~\bibnamefont
  {Wang}}, \bibinfo {author} {\bibfnamefont {D.}~\bibnamefont {Xue}}, \bibinfo
  {author} {\bibfnamefont {H.}~\bibnamefont {Yang}},\ and\ \bibinfo {author}
  {\bibfnamefont {F.}~\bibnamefont {Pan}},\ }\bibfield  {title} {\bibinfo
  {title} {{Observation of the antiferromagnetic spin Hall effect}},\ }\href
  {https://doi.org/10.1038/s41563-021-00946-z} {\bibfield  {journal} {\bibinfo
  {journal} {Nature Materials}\ }\textbf {\bibinfo {volume} {20}},\ \bibinfo
  {pages} {800} (\bibinfo {year} {2021})}\BibitemShut {NoStop}%
\bibitem [{\citenamefont {Bose}\ \emph {et~al.}(2022)\citenamefont {Bose},
  \citenamefont {Schreiber}, \citenamefont {Jain}, \citenamefont {Shao},
  \citenamefont {Nair}, \citenamefont {Sun}, \citenamefont {Zhang},
  \citenamefont {Muller}, \citenamefont {Tsymbal}, \citenamefont {Schlom},\
  and\ \citenamefont {Ralph}}]{Bose2022}%
  \BibitemOpen
  \bibfield  {author} {\bibinfo {author} {\bibfnamefont {A.}~\bibnamefont
  {Bose}}, \bibinfo {author} {\bibfnamefont {N.~J.}\ \bibnamefont {Schreiber}},
  \bibinfo {author} {\bibfnamefont {R.}~\bibnamefont {Jain}}, \bibinfo {author}
  {\bibfnamefont {D.~F.}\ \bibnamefont {Shao}}, \bibinfo {author}
  {\bibfnamefont {H.~P.}\ \bibnamefont {Nair}}, \bibinfo {author}
  {\bibfnamefont {J.}~\bibnamefont {Sun}}, \bibinfo {author} {\bibfnamefont
  {X.~S.}\ \bibnamefont {Zhang}}, \bibinfo {author} {\bibfnamefont {D.~A.}\
  \bibnamefont {Muller}}, \bibinfo {author} {\bibfnamefont {E.~Y.}\
  \bibnamefont {Tsymbal}}, \bibinfo {author} {\bibfnamefont {D.~G.}\
  \bibnamefont {Schlom}},\ and\ \bibinfo {author} {\bibfnamefont {D.~C.}\
  \bibnamefont {Ralph}},\ }\bibfield  {title} {\bibinfo {title} {{Tilted spin
  current generated by the collinear antiferromagnet ruthenium dioxide}},\
  }\href {https://doi.org/10.1038/s41928-022-00744-8} {\bibfield  {journal}
  {\bibinfo  {journal} {Nature Electronics}\ }\textbf {\bibinfo {volume} {5}},\
  \bibinfo {pages} {267} (\bibinfo {year} {2022})}\BibitemShut {NoStop}%
\bibitem [{\citenamefont {Bai}\ \emph {et~al.}(2022)\citenamefont {Bai},
  \citenamefont {Han}, \citenamefont {Feng}, \citenamefont {Zhou},
  \citenamefont {Su}, \citenamefont {Wang}, \citenamefont {Liao}, \citenamefont
  {Zhu}, \citenamefont {Chen}, \citenamefont {Pan}, \citenamefont {Fan},\ and\
  \citenamefont {Song}}]{Bai2022}%
  \BibitemOpen
  \bibfield  {author} {\bibinfo {author} {\bibfnamefont {H.}~\bibnamefont
  {Bai}}, \bibinfo {author} {\bibfnamefont {L.}~\bibnamefont {Han}}, \bibinfo
  {author} {\bibfnamefont {X.~Y.}\ \bibnamefont {Feng}}, \bibinfo {author}
  {\bibfnamefont {Y.~J.}\ \bibnamefont {Zhou}}, \bibinfo {author}
  {\bibfnamefont {R.~X.}\ \bibnamefont {Su}}, \bibinfo {author} {\bibfnamefont
  {Q.}~\bibnamefont {Wang}}, \bibinfo {author} {\bibfnamefont {L.~Y.}\
  \bibnamefont {Liao}}, \bibinfo {author} {\bibfnamefont {W.~X.}\ \bibnamefont
  {Zhu}}, \bibinfo {author} {\bibfnamefont {X.~Z.}\ \bibnamefont {Chen}},
  \bibinfo {author} {\bibfnamefont {F.}~\bibnamefont {Pan}}, \bibinfo {author}
  {\bibfnamefont {X.~L.}\ \bibnamefont {Fan}},\ and\ \bibinfo {author}
  {\bibfnamefont {C.}~\bibnamefont {Song}},\ }\bibfield  {title} {\bibinfo
  {title} {{Observation of Spin Splitting Torque in a Collinear Antiferromagnet
  RuO 2}},\ }\href {https://doi.org/10.1103/physrevlett.128.197202} {\bibfield
  {journal} {\bibinfo  {journal} {Physical Review Letters}\ }\textbf {\bibinfo
  {volume} {128}},\ \bibinfo {pages} {197202} (\bibinfo {year} {2022})},\
  \Eprint {https://arxiv.org/abs/2109.05933} {arXiv:2109.05933} \BibitemShut
  {NoStop}%
\bibitem [{\citenamefont {Nan}\ \emph {et~al.}(2020)\citenamefont {Nan},
  \citenamefont {Quintela}, \citenamefont {Irwin}, \citenamefont {Gurung},
  \citenamefont {Shao}, \citenamefont {Gibbons}, \citenamefont {Campbell},
  \citenamefont {Song}, \citenamefont {Choi}, \citenamefont {Guo},
  \citenamefont {Johnson}, \citenamefont {Manuel}, \citenamefont {Chopdekar},
  \citenamefont {Hallsteinsen}, \citenamefont {Tybell}, \citenamefont {Ryan},
  \citenamefont {Kim}, \citenamefont {Choi}, \citenamefont {Radaelli},
  \citenamefont {Ralph}, \citenamefont {Tsymbal}, \citenamefont {Rzchowski},\
  and\ \citenamefont {Eom}}]{Nan2020}%
  \BibitemOpen
  \bibfield  {author} {\bibinfo {author} {\bibfnamefont {T.}~\bibnamefont
  {Nan}}, \bibinfo {author} {\bibfnamefont {C.~X.}\ \bibnamefont {Quintela}},
  \bibinfo {author} {\bibfnamefont {J.}~\bibnamefont {Irwin}}, \bibinfo
  {author} {\bibfnamefont {G.}~\bibnamefont {Gurung}}, \bibinfo {author}
  {\bibfnamefont {D.~F.}\ \bibnamefont {Shao}}, \bibinfo {author}
  {\bibfnamefont {J.}~\bibnamefont {Gibbons}}, \bibinfo {author} {\bibfnamefont
  {N.}~\bibnamefont {Campbell}}, \bibinfo {author} {\bibfnamefont
  {K.}~\bibnamefont {Song}}, \bibinfo {author} {\bibfnamefont {S.~Y.}\
  \bibnamefont {Choi}}, \bibinfo {author} {\bibfnamefont {L.}~\bibnamefont
  {Guo}}, \bibinfo {author} {\bibfnamefont {R.~D.}\ \bibnamefont {Johnson}},
  \bibinfo {author} {\bibfnamefont {P.}~\bibnamefont {Manuel}}, \bibinfo
  {author} {\bibfnamefont {R.~V.}\ \bibnamefont {Chopdekar}}, \bibinfo {author}
  {\bibfnamefont {I.}~\bibnamefont {Hallsteinsen}}, \bibinfo {author}
  {\bibfnamefont {T.}~\bibnamefont {Tybell}}, \bibinfo {author} {\bibfnamefont
  {P.~J.}\ \bibnamefont {Ryan}}, \bibinfo {author} {\bibfnamefont {J.~W.}\
  \bibnamefont {Kim}}, \bibinfo {author} {\bibfnamefont {Y.}~\bibnamefont
  {Choi}}, \bibinfo {author} {\bibfnamefont {P.~G.}\ \bibnamefont {Radaelli}},
  \bibinfo {author} {\bibfnamefont {D.~C.}\ \bibnamefont {Ralph}}, \bibinfo
  {author} {\bibfnamefont {E.~Y.}\ \bibnamefont {Tsymbal}}, \bibinfo {author}
  {\bibfnamefont {M.~S.}\ \bibnamefont {Rzchowski}},\ and\ \bibinfo {author}
  {\bibfnamefont {C.~B.}\ \bibnamefont {Eom}},\ }\bibfield  {title} {\bibinfo
  {title} {{Controlling spin current polarization through non-collinear
  antiferromagnetism}},\ }\href {https://doi.org/10.1038/s41467-020-17999-4}
  {\bibfield  {journal} {\bibinfo  {journal} {Nature Communications}\ }\textbf
  {\bibinfo {volume} {11}},\ \bibinfo {pages} {4671} (\bibinfo {year}
  {2020})},\ \Eprint {https://arxiv.org/abs/1912.12586} {arXiv:1912.12586}
  \BibitemShut {NoStop}%
\bibitem [{\citenamefont {Bai}\ \emph {et~al.}(2021)\citenamefont {Bai},
  \citenamefont {Zhou}, \citenamefont {Zhang}, \citenamefont {Kong},
  \citenamefont {Liao}, \citenamefont {Feng}, \citenamefont {Chen},
  \citenamefont {You}, \citenamefont {Zhou}, \citenamefont {Han}, \citenamefont
  {Zhu}, \citenamefont {Pan}, \citenamefont {Fan},\ and\ \citenamefont
  {Song}}]{Bai2021}%
  \BibitemOpen
  \bibfield  {author} {\bibinfo {author} {\bibfnamefont {H.}~\bibnamefont
  {Bai}}, \bibinfo {author} {\bibfnamefont {X.~F.}\ \bibnamefont {Zhou}},
  \bibinfo {author} {\bibfnamefont {H.~W.}\ \bibnamefont {Zhang}}, \bibinfo
  {author} {\bibfnamefont {W.~W.}\ \bibnamefont {Kong}}, \bibinfo {author}
  {\bibfnamefont {L.~Y.}\ \bibnamefont {Liao}}, \bibinfo {author}
  {\bibfnamefont {X.~Y.}\ \bibnamefont {Feng}}, \bibinfo {author}
  {\bibfnamefont {X.~Z.}\ \bibnamefont {Chen}}, \bibinfo {author}
  {\bibfnamefont {Y.~F.}\ \bibnamefont {You}}, \bibinfo {author} {\bibfnamefont
  {Y.~J.}\ \bibnamefont {Zhou}}, \bibinfo {author} {\bibfnamefont
  {L.}~\bibnamefont {Han}}, \bibinfo {author} {\bibfnamefont {W.~X.}\
  \bibnamefont {Zhu}}, \bibinfo {author} {\bibfnamefont {F.}~\bibnamefont
  {Pan}}, \bibinfo {author} {\bibfnamefont {X.~L.}\ \bibnamefont {Fan}},\ and\
  \bibinfo {author} {\bibfnamefont {C.}~\bibnamefont {Song}},\ }\bibfield
  {title} {\bibinfo {title} {{Control of spin-orbit torques through magnetic
  symmetry in differently oriented noncollinear antiferromagnetic Mn3Pt}},\
  }\href {https://doi.org/10.1103/PhysRevB.104.104401} {\bibfield  {journal}
  {\bibinfo  {journal} {Physical Review B}\ }\textbf {\bibinfo {volume}
  {104}},\ \bibinfo {pages} {104401} (\bibinfo {year} {2021})}\BibitemShut
  {NoStop}%
\bibitem [{\citenamefont {Kondou}\ \emph {et~al.}(2021)\citenamefont {Kondou},
  \citenamefont {Chen}, \citenamefont {Tomita}, \citenamefont {Ikhlas},
  \citenamefont {Higo}, \citenamefont {MacDonald}, \citenamefont {Nakatsuji},\
  and\ \citenamefont {Otani}}]{Koudou2021}%
  \BibitemOpen
  \bibfield  {author} {\bibinfo {author} {\bibfnamefont {K.}~\bibnamefont
  {Kondou}}, \bibinfo {author} {\bibfnamefont {H.}~\bibnamefont {Chen}},
  \bibinfo {author} {\bibfnamefont {T.}~\bibnamefont {Tomita}}, \bibinfo
  {author} {\bibfnamefont {M.}~\bibnamefont {Ikhlas}}, \bibinfo {author}
  {\bibfnamefont {T.}~\bibnamefont {Higo}}, \bibinfo {author} {\bibfnamefont
  {A.~H.}\ \bibnamefont {MacDonald}}, \bibinfo {author} {\bibfnamefont
  {S.}~\bibnamefont {Nakatsuji}},\ and\ \bibinfo {author} {\bibfnamefont
  {Y.~C.}\ \bibnamefont {Otani}},\ }\bibfield  {title} {\bibinfo {title}
  {{Giant field-like torque by the out-of-plane magnetic spin Hall effect in a
  topological antiferromagnet}},\ }\href
  {https://doi.org/10.1038/s41467-021-26453-y} {\bibfield  {journal} {\bibinfo
  {journal} {Nature Communications}\ }\textbf {\bibinfo {volume} {12}},\
  \bibinfo {pages} {1} (\bibinfo {year} {2021})}\BibitemShut {NoStop}%
\bibitem [{\citenamefont {Liu}\ \emph {et~al.}(2021)\citenamefont {Liu},
  \citenamefont {Zhou}, \citenamefont {Shu}, \citenamefont {Li}, \citenamefont
  {Zhao}, \citenamefont {Lin}, \citenamefont {Deng}, \citenamefont {Xie},
  \citenamefont {Chen}, \citenamefont {Zhou}, \citenamefont {Guo},
  \citenamefont {Wang}, \citenamefont {Yu}, \citenamefont {Shi}, \citenamefont
  {Yang}, \citenamefont {Pennycook}, \citenamefont {Manchon},\ and\
  \citenamefont {Chen}}]{Liu2021}%
  \BibitemOpen
  \bibfield  {author} {\bibinfo {author} {\bibfnamefont {L.}~\bibnamefont
  {Liu}}, \bibinfo {author} {\bibfnamefont {C.}~\bibnamefont {Zhou}}, \bibinfo
  {author} {\bibfnamefont {X.}~\bibnamefont {Shu}}, \bibinfo {author}
  {\bibfnamefont {C.}~\bibnamefont {Li}}, \bibinfo {author} {\bibfnamefont
  {T.}~\bibnamefont {Zhao}}, \bibinfo {author} {\bibfnamefont {W.}~\bibnamefont
  {Lin}}, \bibinfo {author} {\bibfnamefont {J.}~\bibnamefont {Deng}}, \bibinfo
  {author} {\bibfnamefont {Q.}~\bibnamefont {Xie}}, \bibinfo {author}
  {\bibfnamefont {S.}~\bibnamefont {Chen}}, \bibinfo {author} {\bibfnamefont
  {J.}~\bibnamefont {Zhou}}, \bibinfo {author} {\bibfnamefont {R.}~\bibnamefont
  {Guo}}, \bibinfo {author} {\bibfnamefont {H.}~\bibnamefont {Wang}}, \bibinfo
  {author} {\bibfnamefont {J.}~\bibnamefont {Yu}}, \bibinfo {author}
  {\bibfnamefont {S.}~\bibnamefont {Shi}}, \bibinfo {author} {\bibfnamefont
  {P.}~\bibnamefont {Yang}}, \bibinfo {author} {\bibfnamefont {S.}~\bibnamefont
  {Pennycook}}, \bibinfo {author} {\bibfnamefont {A.}~\bibnamefont {Manchon}},\
  and\ \bibinfo {author} {\bibfnamefont {J.}~\bibnamefont {Chen}},\ }\bibfield
  {title} {\bibinfo {title} {{Symmetry-dependent field-free switching of
  perpendicular magnetization}},\ }\href
  {https://doi.org/10.1038/s41565-020-00826-8} {\bibfield  {journal} {\bibinfo
  {journal} {Nature Nanotechnology}\ }\textbf {\bibinfo {volume} {16}},\
  \bibinfo {pages} {227} (\bibinfo {year} {2021})}\BibitemShut {NoStop}%
\bibitem [{\citenamefont {{\v{Z}}elezn{\'{y}}}\ \emph
  {et~al.}(2017)\citenamefont {{\v{Z}}elezn{\'{y}}}, \citenamefont {Gao},
  \citenamefont {Manchon}, \citenamefont {Freimuth}, \citenamefont {Mokrousov},
  \citenamefont {Zemen}, \citenamefont {Ma{\v{s}}ek}, \citenamefont {Sinova},\
  and\ \citenamefont {Jungwirth}}]{Zelezny2017}%
  \BibitemOpen
  \bibfield  {author} {\bibinfo {author} {\bibfnamefont {J.}~\bibnamefont
  {{\v{Z}}elezn{\'{y}}}}, \bibinfo {author} {\bibfnamefont {H.}~\bibnamefont
  {Gao}}, \bibinfo {author} {\bibfnamefont {A.}~\bibnamefont {Manchon}},
  \bibinfo {author} {\bibfnamefont {F.}~\bibnamefont {Freimuth}}, \bibinfo
  {author} {\bibfnamefont {Y.}~\bibnamefont {Mokrousov}}, \bibinfo {author}
  {\bibfnamefont {J.}~\bibnamefont {Zemen}}, \bibinfo {author} {\bibfnamefont
  {J.}~\bibnamefont {Ma{\v{s}}ek}}, \bibinfo {author} {\bibfnamefont
  {J.}~\bibnamefont {Sinova}},\ and\ \bibinfo {author} {\bibfnamefont
  {T.}~\bibnamefont {Jungwirth}},\ }\bibfield  {title} {\bibinfo {title}
  {{Spin-orbit torques in locally and globally non-centrosymmetric crystals:
  Antiferromagnets and ferromagnets}},\ }\href
  {https://doi.org/10.1103/PhysRevB.95.014403} {\bibfield  {journal} {\bibinfo
  {journal} {Physical Review B}\ }\textbf {\bibinfo {volume} {95}},\ \bibinfo
  {pages} {014403} (\bibinfo {year} {2017})}\BibitemShut {NoStop}%
\bibitem [{\citenamefont {Gong}\ and\ \citenamefont {Zhang}(2019)}]{Gong2019}%
  \BibitemOpen
  \bibfield  {author} {\bibinfo {author} {\bibfnamefont {C.}~\bibnamefont
  {Gong}}\ and\ \bibinfo {author} {\bibfnamefont {X.}~\bibnamefont {Zhang}},\
  }\bibfield  {title} {\bibinfo {title} {{Two-dimensional magnetic crystals and
  emergent heterostructure devices}},\ }\href
  {https://doi.org/10.1126/science.aav4450} {\bibfield  {journal} {\bibinfo
  {journal} {Science}\ }\textbf {\bibinfo {volume} {363}},\ \bibinfo {pages}
  {706} (\bibinfo {year} {2019})}\BibitemShut {NoStop}%
\bibitem [{\citenamefont {Johansen}\ \emph {et~al.}(2019)\citenamefont
  {Johansen}, \citenamefont {Risingg{\aa}rd}, \citenamefont {Sudb{\o}},
  \citenamefont {Linder},\ and\ \citenamefont {Brataas}}]{Johansen2019}%
  \BibitemOpen
  \bibfield  {author} {\bibinfo {author} {\bibfnamefont {{\O}.}~\bibnamefont
  {Johansen}}, \bibinfo {author} {\bibfnamefont {V.}~\bibnamefont
  {Risingg{\aa}rd}}, \bibinfo {author} {\bibfnamefont {A.}~\bibnamefont
  {Sudb{\o}}}, \bibinfo {author} {\bibfnamefont {J.}~\bibnamefont {Linder}},\
  and\ \bibinfo {author} {\bibfnamefont {A.}~\bibnamefont {Brataas}},\
  }\bibfield  {title} {\bibinfo {title} {{Current Control of Magnetism in
  Two-Dimensional Fe3GeTe2}},\ }\href
  {https://doi.org/10.1103/PhysRevLett.122.217203} {\bibfield  {journal}
  {\bibinfo  {journal} {Physical Review Letters}\ }\textbf {\bibinfo {volume}
  {122}},\ \bibinfo {pages} {217203} (\bibinfo {year} {2019})}\BibitemShut
  {NoStop}%
\bibitem [{\citenamefont {Zhang}\ \emph {et~al.}(2021)\citenamefont {Zhang},
  \citenamefont {Han}, \citenamefont {Lee}, \citenamefont {Coak}, \citenamefont
  {Kim}, \citenamefont {Hwang}, \citenamefont {Son}, \citenamefont {Shin},
  \citenamefont {Lim}, \citenamefont {Jo}, \citenamefont {Kim}, \citenamefont
  {Kim}, \citenamefont {Lee},\ and\ \citenamefont {Park}}]{Zhang2021}%
  \BibitemOpen
  \bibfield  {author} {\bibinfo {author} {\bibfnamefont {K.}~\bibnamefont
  {Zhang}}, \bibinfo {author} {\bibfnamefont {S.}~\bibnamefont {Han}}, \bibinfo
  {author} {\bibfnamefont {Y.}~\bibnamefont {Lee}}, \bibinfo {author}
  {\bibfnamefont {M.~J.}\ \bibnamefont {Coak}}, \bibinfo {author}
  {\bibfnamefont {J.}~\bibnamefont {Kim}}, \bibinfo {author} {\bibfnamefont
  {I.}~\bibnamefont {Hwang}}, \bibinfo {author} {\bibfnamefont
  {S.}~\bibnamefont {Son}}, \bibinfo {author} {\bibfnamefont {J.}~\bibnamefont
  {Shin}}, \bibinfo {author} {\bibfnamefont {M.}~\bibnamefont {Lim}}, \bibinfo
  {author} {\bibfnamefont {D.}~\bibnamefont {Jo}}, \bibinfo {author}
  {\bibfnamefont {K.}~\bibnamefont {Kim}}, \bibinfo {author} {\bibfnamefont
  {D.}~\bibnamefont {Kim}}, \bibinfo {author} {\bibfnamefont {H.~W.}\
  \bibnamefont {Lee}},\ and\ \bibinfo {author} {\bibfnamefont {J.~G.}\
  \bibnamefont {Park}},\ }\bibfield  {title} {\bibinfo {title} {{Gigantic
  Current Control of Coercive Field and Magnetic Memory Based on Nanometer-Thin
  Ferromagnetic van der Waals Fe3GeTe2}},\ }\href
  {https://doi.org/10.1002/adma.202004110} {\bibfield  {journal} {\bibinfo
  {journal} {Advanced Materials}\ }\textbf {\bibinfo {volume} {33}},\ \bibinfo
  {pages} {2004110} (\bibinfo {year} {2021})}\BibitemShut {NoStop}%
\bibitem [{\citenamefont {Laref}\ \emph {et~al.}(2020)\citenamefont {Laref},
  \citenamefont {Kim},\ and\ \citenamefont {Manchon}}]{Laref2020c}%
  \BibitemOpen
  \bibfield  {author} {\bibinfo {author} {\bibfnamefont {S.}~\bibnamefont
  {Laref}}, \bibinfo {author} {\bibfnamefont {K.-w.}\ \bibnamefont {Kim}},\
  and\ \bibinfo {author} {\bibfnamefont {A.}~\bibnamefont {Manchon}},\
  }\bibfield  {title} {\bibinfo {title} {{Elusive Dzyaloshinskii-Moriya
  interaction in monolayer Fe3GeTe2}},\ }\href
  {https://doi.org/10.1103/PhysRevB.102.060402} {\bibfield  {journal} {\bibinfo
   {journal} {Physical Review B}\ }\textbf {\bibinfo {volume} {102}},\ \bibinfo
  {pages} {060402(R)} (\bibinfo {year} {2020})}\BibitemShut {NoStop}%
\bibitem [{\citenamefont {Smaili}\ \emph {et~al.}(2021)\citenamefont {Smaili},
  \citenamefont {Laref}, \citenamefont {Garcia}, \citenamefont
  {Schwingenschl{\"{o}}gl}, \citenamefont {Roche},\ and\ \citenamefont
  {Manchon}}]{Smaili2021}%
  \BibitemOpen
  \bibfield  {author} {\bibinfo {author} {\bibfnamefont {I.}~\bibnamefont
  {Smaili}}, \bibinfo {author} {\bibfnamefont {S.}~\bibnamefont {Laref}},
  \bibinfo {author} {\bibfnamefont {J.~H.}\ \bibnamefont {Garcia}}, \bibinfo
  {author} {\bibfnamefont {U.}~\bibnamefont {Schwingenschl{\"{o}}gl}}, \bibinfo
  {author} {\bibfnamefont {S.}~\bibnamefont {Roche}},\ and\ \bibinfo {author}
  {\bibfnamefont {A.}~\bibnamefont {Manchon}},\ }\bibfield  {title} {\bibinfo
  {title} {{Janus monolayers of magnetic transition metal dichalcogenides as an
  all-in-one platform for spin-orbit torque}},\ }\href
  {https://doi.org/10.1103/PhysRevB.104.104415} {\bibfield  {journal} {\bibinfo
   {journal} {Physical Review B}\ }\textbf {\bibinfo {volume} {104}},\ \bibinfo
  {pages} {104415} (\bibinfo {year} {2021})},\ \Eprint
  {https://arxiv.org/abs/2007.07579} {arXiv:2007.07579} \BibitemShut {NoStop}%
\bibitem [{\citenamefont {Dresselhaus}(2008)}]{Dresselhaus2008}%
  \BibitemOpen
  \bibfield  {author} {\bibinfo {author} {\bibfnamefont {M.~S.}\ \bibnamefont
  {Dresselhaus}},\ }\href@noop {} {\emph {\bibinfo {title} {{Group Theory:
  Application to the Physics of Condensed Matter}}}}\ (\bibinfo  {publisher}
  {Springer},\ \bibinfo {year} {2008})\BibitemShut {NoStop}%
\bibitem [{\citenamefont {Lax}(2012)}]{Lax2012}%
  \BibitemOpen
  \bibfield  {author} {\bibinfo {author} {\bibfnamefont {M.~J.}\ \bibnamefont
  {Lax}},\ }\href@noop {} {\emph {\bibinfo {title} {{Symmetry Principles in
  Solid State and Molecular Physics}}}}\ (\bibinfo  {publisher} {Dover
  Publications},\ \bibinfo {address} {Dover, UK},\ \bibinfo {year}
  {2012})\BibitemShut {NoStop}%
\bibitem [{\citenamefont {Liu}\ \emph {et~al.}(2012)\citenamefont {Liu},
  \citenamefont {Pai}, \citenamefont {Ralph},\ and\ \citenamefont
  {Buhrman}}]{Liu2012b}%
  \BibitemOpen
  \bibfield  {author} {\bibinfo {author} {\bibfnamefont {L.}~\bibnamefont
  {Liu}}, \bibinfo {author} {\bibfnamefont {C.-F.}\ \bibnamefont {Pai}},
  \bibinfo {author} {\bibfnamefont {D.~C.}\ \bibnamefont {Ralph}},\ and\
  \bibinfo {author} {\bibfnamefont {R.~a.}\ \bibnamefont {Buhrman}},\
  }\bibfield  {title} {\bibinfo {title} {{Magnetic Oscillations Driven by the
  Spin Hall Effect in 3-Terminal Magnetic Tunnel Junction Devices}},\ }\href
  {https://doi.org/10.1103/PhysRevLett.109.186602} {\bibfield  {journal}
  {\bibinfo  {journal} {Physical Review Letters}\ }\textbf {\bibinfo {volume}
  {109}},\ \bibinfo {pages} {186602} (\bibinfo {year} {2012})}\BibitemShut
  {NoStop}%
\bibitem [{\citenamefont {Demidov}\ \emph {et~al.}(2012)\citenamefont
  {Demidov}, \citenamefont {Urazhdin}, \citenamefont {Ulrichs}, \citenamefont
  {Tiberkevich}, \citenamefont {Slavin}, \citenamefont {Baither}, \citenamefont
  {Schmitz},\ and\ \citenamefont {Demokritov}}]{Demidov2012}%
  \BibitemOpen
  \bibfield  {author} {\bibinfo {author} {\bibfnamefont {V.~E.}\ \bibnamefont
  {Demidov}}, \bibinfo {author} {\bibfnamefont {S.}~\bibnamefont {Urazhdin}},
  \bibinfo {author} {\bibfnamefont {H.}~\bibnamefont {Ulrichs}}, \bibinfo
  {author} {\bibfnamefont {V.}~\bibnamefont {Tiberkevich}}, \bibinfo {author}
  {\bibfnamefont {A.}~\bibnamefont {Slavin}}, \bibinfo {author} {\bibfnamefont
  {D.}~\bibnamefont {Baither}}, \bibinfo {author} {\bibfnamefont
  {G.}~\bibnamefont {Schmitz}},\ and\ \bibinfo {author} {\bibfnamefont {S.~O.}\
  \bibnamefont {Demokritov}},\ }\bibfield  {title} {\bibinfo {title} {{Magnetic
  nano-oscillator driven by pure spin current.}},\ }\href
  {https://doi.org/10.1038/nmat3459} {\bibfield  {journal} {\bibinfo  {journal}
  {Nature Materials}\ }\textbf {\bibinfo {volume} {11}},\ \bibinfo {pages}
  {1028} (\bibinfo {year} {2012})}\BibitemShut {NoStop}%
\bibitem [{\citenamefont {Kurebayashi}\ and\ \citenamefont
  {Nagaosa}(2019)}]{Kurebayashi2019}%
  \BibitemOpen
  \bibfield  {author} {\bibinfo {author} {\bibfnamefont {D.}~\bibnamefont
  {Kurebayashi}}\ and\ \bibinfo {author} {\bibfnamefont {N.}~\bibnamefont
  {Nagaosa}},\ }\bibfield  {title} {\bibinfo {title} {{Theory of current-driven
  dynamics of spin textures on the surface of a topological insulator}},\
  }\href {https://doi.org/10.1103/PhysRevB.100.134407} {\bibfield  {journal}
  {\bibinfo  {journal} {Physical Review B}\ }\textbf {\bibinfo {volume}
  {100}},\ \bibinfo {pages} {134407} (\bibinfo {year} {2019})},\ \Eprint
  {https://arxiv.org/abs/1908.00152} {arXiv:1908.00152} \BibitemShut {NoStop}%
\bibitem [{\citenamefont {Imai}\ \emph {et~al.}(2021)\citenamefont {Imai},
  \citenamefont {Yamaguchi}, \citenamefont {Yamakage},\ and\ \citenamefont
  {Kohno}}]{Imai2021}%
  \BibitemOpen
  \bibfield  {author} {\bibinfo {author} {\bibfnamefont {Y.}~\bibnamefont
  {Imai}}, \bibinfo {author} {\bibfnamefont {T.}~\bibnamefont {Yamaguchi}},
  \bibinfo {author} {\bibfnamefont {A.}~\bibnamefont {Yamakage}},\ and\
  \bibinfo {author} {\bibfnamefont {H.}~\bibnamefont {Kohno}},\ }\bibfield
  {title} {\bibinfo {title} {{Spintronic properties of topological surface
  Dirac electrons with hexagonal warping}},\ }\href
  {https://doi.org/10.1103/PhysRevB.103.054402} {\bibfield  {journal} {\bibinfo
   {journal} {Physical Review B}\ }\textbf {\bibinfo {volume} {103}},\ \bibinfo
  {pages} {54402} (\bibinfo {year} {2021})}\BibitemShut {NoStop}%
\bibitem [{\citenamefont {Zhou}\ \emph {et~al.}(2022)\citenamefont {Zhou},
  \citenamefont {Duan}, \citenamefont {Wu}, \citenamefont {Deng}, \citenamefont
  {Wang}, \citenamefont {Culcer},\ and\ \citenamefont {Wang}}]{Zhou2022}%
  \BibitemOpen
  \bibfield  {author} {\bibinfo {author} {\bibfnamefont {Y.~L.}\ \bibnamefont
  {Zhou}}, \bibinfo {author} {\bibfnamefont {H.~J.}\ \bibnamefont {Duan}},
  \bibinfo {author} {\bibfnamefont {Y.~J.}\ \bibnamefont {Wu}}, \bibinfo
  {author} {\bibfnamefont {M.~X.}\ \bibnamefont {Deng}}, \bibinfo {author}
  {\bibfnamefont {L.}~\bibnamefont {Wang}}, \bibinfo {author} {\bibfnamefont
  {D.}~\bibnamefont {Culcer}},\ and\ \bibinfo {author} {\bibfnamefont {R.~Q.}\
  \bibnamefont {Wang}},\ }\bibfield  {title} {\bibinfo {title} {{Nonlinear
  antidamping spin-orbit torque originating from intraband transport on the
  warped surface of a topological insulator}},\ }\href
  {https://doi.org/10.1103/PhysRevB.105.075415} {\bibfield  {journal} {\bibinfo
   {journal} {Physical Review B}\ }\textbf {\bibinfo {volume} {105}},\ \bibinfo
  {pages} {1} (\bibinfo {year} {2022})},\ \Eprint
  {https://arxiv.org/abs/2111.03397} {arXiv:2111.03397} \BibitemShut {NoStop}%
\bibitem [{\citenamefont {Li}\ \emph {et~al.}(2019{\natexlab{b}})\citenamefont
  {Li}, \citenamefont {Wang}, \citenamefont {Deng},\ and\ \citenamefont
  {Yang}}]{Li2019k}%
  \BibitemOpen
  \bibfield  {author} {\bibinfo {author} {\bibfnamefont {J.~Y.}\ \bibnamefont
  {Li}}, \bibinfo {author} {\bibfnamefont {R.~Q.}\ \bibnamefont {Wang}},
  \bibinfo {author} {\bibfnamefont {M.~X.}\ \bibnamefont {Deng}},\ and\
  \bibinfo {author} {\bibfnamefont {M.}~\bibnamefont {Yang}},\ }\bibfield
  {title} {\bibinfo {title} {{In-plane magnetization effect on current-induced
  spin-orbit torque in a ferromagnet/topological insulator bilayer with
  hexagonal warping}},\ }\href {https://doi.org/10.1103/PhysRevB.99.155139}
  {\bibfield  {journal} {\bibinfo  {journal} {Physical Review B}\ }\textbf
  {\bibinfo {volume} {99}},\ \bibinfo {pages} {155139} (\bibinfo {year}
  {2019}{\natexlab{b}})}\BibitemShut {NoStop}%
\bibitem [{\citenamefont {Bonbien}\ and\ \citenamefont
  {Manchon}(2020)}]{Bonbien2020}%
  \BibitemOpen
  \bibfield  {author} {\bibinfo {author} {\bibfnamefont {V.}~\bibnamefont
  {Bonbien}}\ and\ \bibinfo {author} {\bibfnamefont {A.}~\bibnamefont
  {Manchon}},\ }\bibfield  {title} {\bibinfo {title} {{Symmetrized
  decomposition of the Kubo-Bastin formula}},\ }\href
  {https://doi.org/10.1103/PhysRevB.102.085113} {\bibfield  {journal} {\bibinfo
   {journal} {Physical Review B}\ }\textbf {\bibinfo {volume} {102}},\ \bibinfo
  {pages} {085113} (\bibinfo {year} {2020})}\BibitemShut {NoStop}%
\bibitem [{\citenamefont {Chen}\ \emph {et~al.}(2009)\citenamefont {Chen},
  \citenamefont {Analytis}, \citenamefont {Chu}, \citenamefont {Liu},
  \citenamefont {Mo}, \citenamefont {Qi}, \citenamefont {Zhang}, \citenamefont
  {Lu}, \citenamefont {Dai}, \citenamefont {Fang}, \citenamefont {Zhang},
  \citenamefont {Fisher}, \citenamefont {Hussain},\ and\ \citenamefont
  {Shen}}]{Chen2009}%
  \BibitemOpen
  \bibfield  {author} {\bibinfo {author} {\bibfnamefont {Y.~L.}\ \bibnamefont
  {Chen}}, \bibinfo {author} {\bibfnamefont {J.~G.}\ \bibnamefont {Analytis}},
  \bibinfo {author} {\bibfnamefont {J.-H.}\ \bibnamefont {Chu}}, \bibinfo
  {author} {\bibfnamefont {Z.~K.}\ \bibnamefont {Liu}}, \bibinfo {author}
  {\bibfnamefont {S.-K.}\ \bibnamefont {Mo}}, \bibinfo {author} {\bibfnamefont
  {X.~L.}\ \bibnamefont {Qi}}, \bibinfo {author} {\bibfnamefont {H.~J.}\
  \bibnamefont {Zhang}}, \bibinfo {author} {\bibfnamefont {D.~H.}\ \bibnamefont
  {Lu}}, \bibinfo {author} {\bibfnamefont {X.}~\bibnamefont {Dai}}, \bibinfo
  {author} {\bibfnamefont {Z.}~\bibnamefont {Fang}}, \bibinfo {author}
  {\bibfnamefont {S.~C.}\ \bibnamefont {Zhang}}, \bibinfo {author}
  {\bibfnamefont {I.~R.}\ \bibnamefont {Fisher}}, \bibinfo {author}
  {\bibfnamefont {Z.}~\bibnamefont {Hussain}},\ and\ \bibinfo {author}
  {\bibfnamefont {Z.-X.}\ \bibnamefont {Shen}},\ }\bibfield  {title} {\bibinfo
  {title} {{Experimental realization of a three-dimensional topological
  insulator, Bi2Te3.}},\ }\href {https://doi.org/10.1126/science.1173034}
  {\bibfield  {journal} {\bibinfo  {journal} {Science (New York, N.Y.)}\
  }\textbf {\bibinfo {volume} {325}},\ \bibinfo {pages} {178} (\bibinfo {year}
  {2009})}\BibitemShut {NoStop}%
\bibitem [{\citenamefont {Hsieh}\ \emph {et~al.}(2009)\citenamefont {Hsieh},
  \citenamefont {Xia}, \citenamefont {Qian}, \citenamefont {Wray},
  \citenamefont {Dil}, \citenamefont {Meier}, \citenamefont {Osterwalder},
  \citenamefont {Patthey}, \citenamefont {Checkelsky}, \citenamefont {Ong},
  \citenamefont {Fedorov}, \citenamefont {Lin}, \citenamefont {Bansil},
  \citenamefont {Grauer}, \citenamefont {Hor}, \citenamefont {Cava},\ and\
  \citenamefont {Hasan}}]{Hsieh2009}%
  \BibitemOpen
  \bibfield  {author} {\bibinfo {author} {\bibfnamefont {D.}~\bibnamefont
  {Hsieh}}, \bibinfo {author} {\bibfnamefont {Y.}~\bibnamefont {Xia}}, \bibinfo
  {author} {\bibfnamefont {D.}~\bibnamefont {Qian}}, \bibinfo {author}
  {\bibfnamefont {L.}~\bibnamefont {Wray}}, \bibinfo {author} {\bibfnamefont
  {J.~H.}\ \bibnamefont {Dil}}, \bibinfo {author} {\bibfnamefont
  {F.}~\bibnamefont {Meier}}, \bibinfo {author} {\bibfnamefont
  {J.}~\bibnamefont {Osterwalder}}, \bibinfo {author} {\bibfnamefont
  {L.}~\bibnamefont {Patthey}}, \bibinfo {author} {\bibfnamefont {J.~G.}\
  \bibnamefont {Checkelsky}}, \bibinfo {author} {\bibfnamefont {N.~P.}\
  \bibnamefont {Ong}}, \bibinfo {author} {\bibfnamefont {A.~V.}\ \bibnamefont
  {Fedorov}}, \bibinfo {author} {\bibfnamefont {H.}~\bibnamefont {Lin}},
  \bibinfo {author} {\bibfnamefont {A.}~\bibnamefont {Bansil}}, \bibinfo
  {author} {\bibfnamefont {D.}~\bibnamefont {Grauer}}, \bibinfo {author}
  {\bibfnamefont {Y.~S.}\ \bibnamefont {Hor}}, \bibinfo {author} {\bibfnamefont
  {R.~J.}\ \bibnamefont {Cava}},\ and\ \bibinfo {author} {\bibfnamefont
  {M.~Z.}\ \bibnamefont {Hasan}},\ }\bibfield  {title} {\bibinfo {title} {{A
  tunable topological insulator in the spin helical Dirac transport regime}},\
  }\href {http://dx.doi.org/10.1038/nature08234
  http://www.nature.com/nature/journal/v460/n7259/suppinfo/nature08234_S1.html}
  {\bibfield  {journal} {\bibinfo  {journal} {Nature}\ }\textbf {\bibinfo
  {volume} {460}},\ \bibinfo {pages} {1101} (\bibinfo {year}
  {2009})}\BibitemShut {NoStop}%
\bibitem [{\citenamefont {Alpichshev}\ \emph {et~al.}(2010)\citenamefont
  {Alpichshev}, \citenamefont {Analytis}, \citenamefont {Chu}, \citenamefont
  {Fisher}, \citenamefont {Chen}, \citenamefont {Shen}, \citenamefont {Fang},\
  and\ \citenamefont {Kapitulnik}}]{Alpichshev2010}%
  \BibitemOpen
  \bibfield  {author} {\bibinfo {author} {\bibfnamefont {Z.}~\bibnamefont
  {Alpichshev}}, \bibinfo {author} {\bibfnamefont {J.~G.}\ \bibnamefont
  {Analytis}}, \bibinfo {author} {\bibfnamefont {J.~H.}\ \bibnamefont {Chu}},
  \bibinfo {author} {\bibfnamefont {I.~R.}\ \bibnamefont {Fisher}}, \bibinfo
  {author} {\bibfnamefont {Y.~L.}\ \bibnamefont {Chen}}, \bibinfo {author}
  {\bibfnamefont {Z.~X.}\ \bibnamefont {Shen}}, \bibinfo {author}
  {\bibfnamefont {A.}~\bibnamefont {Fang}},\ and\ \bibinfo {author}
  {\bibfnamefont {A.}~\bibnamefont {Kapitulnik}},\ }\bibfield  {title}
  {\bibinfo {title} {{STM imaging of electronic waves on the surface of Bi2Te3:
  Topologically protected surface states and hexagonal warping effects}},\
  }\href {https://doi.org/10.1103/PhysRevLett.104.016401} {\bibfield  {journal}
  {\bibinfo  {journal} {Physical Review Letters}\ }\textbf {\bibinfo {volume}
  {104}},\ \bibinfo {pages} {016401} (\bibinfo {year} {2010})},\ \Eprint
  {https://arxiv.org/abs/0908.0371} {arXiv:0908.0371} \BibitemShut {NoStop}%
\bibitem [{\citenamefont {Chagas}\ \emph {et~al.}(2022)\citenamefont {Chagas},
  \citenamefont {Ashour}, \citenamefont {Ribeiro}, \citenamefont {Silva},
  \citenamefont {Li}, \citenamefont {Louie}, \citenamefont
  {Magalh{\~{a}}es-Paniago},\ and\ \citenamefont {Petroff}}]{Chagas2022}%
  \BibitemOpen
  \bibfield  {author} {\bibinfo {author} {\bibfnamefont {T.}~\bibnamefont
  {Chagas}}, \bibinfo {author} {\bibfnamefont {O.~A.}\ \bibnamefont {Ashour}},
  \bibinfo {author} {\bibfnamefont {G.~A.~S.}\ \bibnamefont {Ribeiro}},
  \bibinfo {author} {\bibfnamefont {W.~S.}\ \bibnamefont {Silva}}, \bibinfo
  {author} {\bibfnamefont {Z.}~\bibnamefont {Li}}, \bibinfo {author}
  {\bibfnamefont {S.~G.}\ \bibnamefont {Louie}}, \bibinfo {author}
  {\bibfnamefont {R.}~\bibnamefont {Magalh{\~{a}}es-Paniago}},\ and\ \bibinfo
  {author} {\bibfnamefont {Y.}~\bibnamefont {Petroff}},\ }\bibfield  {title}
  {\bibinfo {title} {{Multiple strong topological gaps and hexagonal warping in
  Bi4Te3}},\ }\href@noop {} {\bibfield  {journal} {\bibinfo  {journal}
  {Physical Review B}\ }\textbf {\bibinfo {volume} {105}},\ \bibinfo {pages}
  {L081409} (\bibinfo {year} {2022})}\BibitemShut {NoStop}%
\bibitem [{\citenamefont {Chen}\ \emph {et~al.}(2022)\citenamefont {Chen},
  \citenamefont {D'Antuono}, \citenamefont {Brookes}, \citenamefont {{De
  Luca}}, \citenamefont {{Di Capua}}, \citenamefont {{Di Gennaro}},
  \citenamefont {Ghiringhelli}, \citenamefont {Piamonteze}, \citenamefont
  {Preziosi}, \citenamefont {Jouault}, \citenamefont {Cabero}, \citenamefont
  {Gonz{\'{a}}lez-Calbet}, \citenamefont {Le{\'{o}}n}, \citenamefont
  {Santamaria}, \citenamefont {Sambri}, \citenamefont {Stornaiuolo},\ and\
  \citenamefont {Salluzzo}}]{Chen2022}%
  \BibitemOpen
  \bibfield  {author} {\bibinfo {author} {\bibfnamefont {Y.}~\bibnamefont
  {Chen}}, \bibinfo {author} {\bibfnamefont {M.}~\bibnamefont {D'Antuono}},
  \bibinfo {author} {\bibfnamefont {N.~B.}\ \bibnamefont {Brookes}}, \bibinfo
  {author} {\bibfnamefont {G.~M.}\ \bibnamefont {{De Luca}}}, \bibinfo {author}
  {\bibfnamefont {R.}~\bibnamefont {{Di Capua}}}, \bibinfo {author}
  {\bibfnamefont {E.}~\bibnamefont {{Di Gennaro}}}, \bibinfo {author}
  {\bibfnamefont {G.}~\bibnamefont {Ghiringhelli}}, \bibinfo {author}
  {\bibfnamefont {C.}~\bibnamefont {Piamonteze}}, \bibinfo {author}
  {\bibfnamefont {D.}~\bibnamefont {Preziosi}}, \bibinfo {author}
  {\bibfnamefont {B.}~\bibnamefont {Jouault}}, \bibinfo {author} {\bibfnamefont
  {M.}~\bibnamefont {Cabero}}, \bibinfo {author} {\bibfnamefont {J.~M.}\
  \bibnamefont {Gonz{\'{a}}lez-Calbet}}, \bibinfo {author} {\bibfnamefont
  {C.}~\bibnamefont {Le{\'{o}}n}}, \bibinfo {author} {\bibfnamefont
  {J.}~\bibnamefont {Santamaria}}, \bibinfo {author} {\bibfnamefont
  {A.}~\bibnamefont {Sambri}}, \bibinfo {author} {\bibfnamefont
  {D.}~\bibnamefont {Stornaiuolo}},\ and\ \bibinfo {author} {\bibfnamefont
  {M.}~\bibnamefont {Salluzzo}},\ }\bibfield  {title} {\bibinfo {title}
  {{Ferromagnetic Quasi-Two-Dimensional Electron Gas with Trigonal Crystal
  Field Splitting}},\ }\href {https://doi.org/10.1021/acsaelm.2c00447}
  {\bibfield  {journal} {\bibinfo  {journal} {ACS Applied Electronic
  Materials}\ }\textbf {\bibinfo {volume} {4}},\ \bibinfo {pages} {3226}
  (\bibinfo {year} {2022})}\BibitemShut {NoStop}%
\end{thebibliography}%

\end{document}